\documentclass[11pt,a4paper]{article}
\usepackage{jheppub}
\usepackage{amsmath}
\usepackage{amssymb}
\usepackage{hyperref}
\usepackage{mathrsfs}
\usepackage{braket}
\usepackage{dsfont}

\newcommand{\heq}{~\hat{=}~}
\newcommand{\lie}{\mathcal{L}}

\title{The Quantum Null Energy Condition in Curved Space}

\author[a]{Zicao Fu}
\author[b,c]{Jason Koeller}
\author[a]{Donald Marolf}

\affiliation[a]{Department of Physics, University of California, Santa Barbara, CA 93106, USA}
\affiliation[b]{Center for Theoretical Physics and Department of Physics,\\
University of California, Berkeley, CA 94720, USA}
\affiliation[c]{Lawrence Berkeley National Laboratory, Berkeley, CA 94720, USA}

\emailAdd{zicaofu@physics.ucsb.edu}
\emailAdd{jkoeller@berkeley.edu}
\emailAdd{marolf@physics.ucsb.edu}

\abstract{
The quantum null energy condition (QNEC) is a conjectured bound on components $(T_{kk} = T_{ab} k^a k^b$) of the stress tensor along a null vector $k^a$ at a point $p$ in terms of a second $k$-derivative of the von Neumann entropy $S$ on one side of a null congruence $N$ through $p$ generated by $k^a$. The conjecture has been established for super-renormalizeable field theories at points $p$ that lie on a bifurcate Killing horizon with null tangent $k^a$ and for large-N holographic theories on flat space. While the Koeller-Leichenauer holographic argument clearly yields an inequality for general $(p,k^a)$, more conditions are generally required for this inequality to be a useful QNEC. For $d\le 3$, for arbitrary backgroud metric we show that the QNEC is naturally finite and independent of renormalization scheme when the expansion $\theta$ of $N$ at the point $p$ vanishes.   This is consistent with the original QNEC conjecture which required $\theta$ and the shear $\sigma_{ab}$ to satisfy $\theta |_p= \dot{\theta}|_p =0$, $\sigma_{ab}|_p=0$.  But for $d=4,5$ more conditions than even these are required. In particular, we also require the vanishing of additional derivatives and a dominant energy condition. In the above cases the holographic argument does indeed yield a finite QNEC, though for $d\ge6$ we argue these properties to fail even for weakly isolated horizons (where all derivatives of $\theta, \sigma_{ab}$ vanish) that also satisfy a dominant energy condition. On the positive side, a corrollary to our work is that, when coupled to Einstein-Hilbert gravity, $d \le 3$ holographic theories at large $N$ satisfy the generalized second law (GSL) of thermodynamics at leading order in Newton's constant $G$. This is the first GSL proof which does not require the quantum fields to be perturbations to a Killing horizon.
}

\begin{document}
\maketitle

\section{Introduction and Summary}

Energy conditions are indispensable in understanding classical and quantum gravity. The weakest but most commonly used of the standard energy conditions is the null energy condition (NEC), which states that \(T_{kk} \equiv T_{ab}k^{a}k^{b} \geq 0\) where \(T_{ab}\) is the stress tensor of the matter coupled to gravity and \(k^{a}\) is any null vector. It is sufficiently weak to be satisfied by familiar classical field theories, yet strong enough to prove the second law of black hole thermodynamics \cite{Hawking:1971tu},  singularity theorems \cite{Penrose:1964wq,HawEll}, the chronology protection theorem \cite{Hawking:1991nk}, topological censorship \cite{Friedman:1993ty}, and other fundamental results.   It also guarantees essential properties of holographic entanglement entropy \cite{Wall:2012uf,Headrick:2014cta} in the context of gauge/gravity duality.

On the other hand, it has long been known that the NEC is violated even in free quantum field theories \cite{Epstein:1965zza}. Several quantum replacements for the NEC have been suggested --- such as the averaged null energy condition (ANEC) \cite{Wald:1991xn,Graham:2007va,Kontou:2012ve,Kontou:2015yha,Kelly:2014mra} and ``quantum inequalities'' \cite{Ford:1994bj,Ford:1996er,Ford:1999qv,Levine:2016bpj} --- which involve integrating \(\braket{T_{kk}}\) over a region of spacetime, and others \cite{Martin-Moruno:2013sfa,Martin-Moruno:2013wfa,Martin-Moruno:2015ena,Martin-Moruno:2017exc}. In this paper we study the quantum null energy condition (QNEC) \cite{Bousso:2015mna,Bousso:2015wca,Koeller:2015qmn,Akers:2016ugt},
\begin{align}\label{QNECintro}
	\braket{T_{kk}} \geq \frac{1}{2\pi} S''~,
\end{align}
which places a bound on the renormalized \(\braket{T_{kk}}\) at a point \(p\) in terms of a particular second derivative of the renormalized von Neumann entropy of a region touching \(p\) with respect to deformations of the region at \(p\).\footnote{The same inequality has been investigated \cite{Fu:2016avb} with the ``causal holographic information'' of \cite{Hubeny:2012wa,Freivogel:2013zta,Hubeny:2013hz,Kelly:2013aja,Bunting:2015sfa} playing the role of \(S\) instead of the von Neumann entropy. Another variant was studied in the hydrodynamic approximation in \cite{Flanagan:1999jp}.} In \eqref{QNECintro} we have set $\hbar =1$.  Although shown in \eqref{QNECintro},  we will often omit the expectation value brackets below.

The conjecture of \cite{Bousso:2015mna} states
that \eqref{QNECintro} holds when $k^a$ generates a locally stationary horizon through $p$; i.e., it generates a hypersurface-orthogonal null congruence with vanishing shear $\sigma_{ab}$ at $p$ and  expansion  $\theta$ at $p$ vanishing to second order  along the generator  ($\sigma_{ab}|_p = \theta|_p = \dot{\theta}|_p = 0 $). Below, we restrict to backgrounds satisfying the null convergence condition $R_{ab}k^a k^b \ge 0$, so the Raychaudhuri equation
\begin{equation}
\label{eq:Ray}
\dot{\theta} = - \frac{\theta^2}{d-2} - \sigma^{ab} \sigma_{ab} - R_{ab} k^a k^b
\end{equation}
then requires $R_{ab}k^a k^b = 0$ at $p$.

This conjecture was
motivated in \cite{Bousso:2015mna} by taking a non-gravitating ($G \rightarrow 0$) limit of a ``quantum focussing conjecture" (QFC), which was in turn motivated by the
generalized second law (GSL) of thermodynamics \cite{Bekenstein:1973ur} and the proposed covariant entropy bound \cite{Bousso:1999xy}. Such conjectures suffice to preserve the most fundamental of the above results even in the presence of quantum corrections  \cite{C:2013uza,Bousso:2015mna}.  For example, although quantum corrections allow the formation of traversable wormholes, the GSL severely limits their utility \cite{Gao:2016bin}.

The QNEC has been proven for deformations along bifurcate Killing horizons in free bosonic theories \cite{Bousso:2015wca} using the techniques of null quantization, and it was also shown to hold for holographic theories formulated in flat space in \cite{Koeller:2015qmn}.   In the holographic case, the Ryu-Takayanagi-Hubeny-Rangamani formula \cite{Hubeny:2007xt} was used to translate the QNEC (applied in the boundary theory) at leading order in \(1/N\) into a statement about how boundary-anchored extremal surfaces in AdS move when the anchoring region is deformed. The relevant condition is that when the boundary region is deformed within its domain of dependence, the corresponding extremal surface should move in a spacelike way, at least near the boundary. This condition --- called ``entanglement wedge nesting'' in \cite{Akers:2016ugt}  --- is automatically true assuming entanglement wedge reconstruction \cite{Czech:2012bh,Headrick:2014cta}, and can also be proven directly from the NEC applied in the bulk \cite{Wall:2012uf}. It was shown in \cite{Akers:2016ugt} that the QNEC continues to hold at all orders in \(1/N\), assuming the entanglement wedge nesting property and the quantum extremal surfaces prescription of \cite{Engelhardt:2014gca} (building on \cite{Faulkner:2013ana}).

And as we note in section \ref{sec:HologProofs} below, the Koeller-Leichenauer holographic argument \cite{Koeller:2015qmn} admits a straightforward extension to arbitrary backgrounds.   However, the quantities to which the resulting inequality applies are naturally divergent.  One might expect that the inequality takes the form of a QNEC for the ``bare'' quantities that have not been fully renormalized. Though even this remains to be shown, it is nevertheless of central interest to understand when local counter-terms contribute to \eqref{QNECintro} -- or, more specifically, when they contribute to the difference between the left- and right-hand sides.  In such cases, a QNEC for renormalized quantities would also depend on the choice of renormalization scheme, as such choices induce finite shifts in naively-divergent couplings.  We therefore refer to this phenomenon as scheme-dependence.

A result of \cite{Wall:2015raa} shows the QNEC to be scheme-independent when the null congruence $N$ lies on a bifurcate Killing horizon (see \cite{Kolekar:2012tq} and \cite{Sarkar:2013swa} for precursors in special cases), but the more general setting is studied in section \ref{sec:schemedep} below.  For $d\le 3$ we find that this extends\footnote{\label{foot:semantic} A semantic subtlety is that \cite{Bousso:2015mna} did not spell out in detail the set of backgrounds in which the QNEC should hold.  In particular, although the focussing theorem applies only to spacetimes satisfying the null convergence condition $R_{ab} k^a k^b \ge 0$ for null vectors $k^a$, the utility of this theorem in Einstein-Hilbert gravity (where $R_{ab} k^a k^b  = 8 \pi T_{ab} k^a k^b$) stems from the fact that reasonable matter theories satisfy the null energy condition (NEC) $T_{ab} k^a k^b \ge 0$.  Thus the focussing theorem holds on solutions to reasonable theories.  The derivation \cite{Bousso:2015mna} of the QNEC from the QFC suggests the former to hold on backgrounds that solve reasonable theories of gravity and indeed the discussion in \cite{Bousso:2015mna} assumed that the QNEC was to be studied on an Einstein space. In contrast, the idea that the QNEC is an intrinsically field-theoretic property (having nothing to do with coupling to gravity) suggests that -- like the classical NEC for reasonable matter theories -- it should in fact hold on {\it any} background spacetime.  We will focus on the latter perspective for several reasons: First, the results of \cite{Bousso:2015wca,Koeller:2015qmn} hold on any bifurcate Killing horizon without other restrictions on the background. Second, as explained in footnote \ref{Fufoot} below, the results of \cite{Fu:2017lps} imply for $d \ge 5$ that the QNEC is scheme-dependent on general null congruences even for standard scalar field theories on Ricci-flat backgrounds. Third, the result below that the QNEC holds on any null congruence $N$ in an arbitrary $d\le 3$ background.} to locally stationary null congruences $N$, and for $d=4,5$ it holds when some additional derivatives of the shear $\sigma_{ab}$ and expansion $\theta$ also vanish at $p$.  For $d \ge 6$ we show that scheme-independence generally fails even for weakly isolated horizons \cite{Ashtekar:2001jb} satisfying the dominant energy condition. The qualitative difference between the above cases is that in \(d\leq5\), finite counter-terms can only depend algebraically on the Riemann tensor by dimensional analysis, while in \(d\geq6\) derivatives of the Riemann tensor become allowed. In Sec.~\ref{scheme-indep} we show explicitly that the QNEC fails to be invariant under the addition of the simplest possible such counter-term, $\int d^{6}x \sqrt{-g}(\nabla_{a}R)(\nabla^{a}R)$.

In all cases where we find the QNEC to be scheme-independent, we show in section \ref{sec:HologProofs} that a renormalized QNEC can be proven for the universal sector of holographic theories using the method of \cite{Koeller:2015qmn}.  This in particular establishes the QNEC for holographic theories on arbitrary backgrounds when the null congruence $N$ lies on a bifurcate Killing horizon.

We close with some final discussion in section \ref{sec:discussion}.  In particular, we note that a corollary to our work is a general proof of the QFC and GSL for holographic quantum field theories in $d\le 3$ at leading order in both $1/N$ and the coupling $G$ to gravity\footnote{I.e., this is the ``boundary'' Newton constant, not the bulk Newton constant of the gravitational dual.}, and for these theories on weakly-isolated horizons in \(d\leq5\). This is the first proof of the GSL in the semi-classical regime which does not require the quantum fields to be perturbations to a Killing horizon.

\section{Scheme-(in)dependence of the QNEC}
\label{sec:schemedep}

As discussed above, a crucial question is whether local counter-terms affect \eqref{QNECintro}.  To answer this question, it is useful to be more precise about how the various terms in \eqref{QNECintro} are to be computed. We briefly review such recipes in section \ref{subsec:prelim} and then consider the effect of local counter-terms in \ref{subsec:dotheycancel}.

\subsection{Preliminaries}
\label{subsec:prelim}

We consider states $|\Psi\rangle$ that are pure on a sufficiently enlarged spacetime (with metric $g_{ab}$) and which are defined by a path integral with arbitrary operator insertions ${\cal O}[\Phi]$ and sources (included in the action $I$):
\begin{align}
\label{PSIPI}
	\ket{\Psi} = \int_{\tau = -\infty}^{\tau = 0} [D\Phi] \mathcal{O}[\Phi] e^{- I[\Phi,g_{ab}]}
\end{align}
We use Euclidean notation for familiarity, though the integration contour may also include Lorentzian or complex pieces of spacetime (as in e.g. the discussion of entropy in \cite{Dong:2016hjy}).

The partition function $Z=\text{Tr}\left[ \left| \Psi  \right\rangle \left\langle  \Psi  \right| \right]$ for such a state is computed by sewing together two copies of \eqref{PSIPI} to form the path integral
\begin{align}
\label{eq:Z}
	Z[g_{ab}] = \int [D\Phi] \mathcal{O}[\Phi] e^{-I[\Phi,g_{ab}]}.
\end{align}
As usual, \eqref{eq:Z} is a functional of the background geometry \(g_{ab}\).  We take the definitions of $[D\Phi]$ and $I[\Phi,g_{ab}]$ to include appropriate renormalizations to make $Z[g_{ab}]$ well-defined. The renormalized effective action $W$ is defined by
\begin{align}
	Z[g_{ab}] = e^{-W[g_{ab}]}.
\end{align}

Both \(T_{kk}\) and \(S''\) are to be computed from \(W[g_{ab}]\). The expectation value of the renormalized gravitational stress tensor in \(\ket{\Psi}\) is defined by
\begin{align}\label{Tdef}
	\braket{T_{ab}} \equiv -\frac{2}{\sqrt{-g}} \frac{\delta W}{\delta g^{ab}}.
\end{align}
Given a region ${\cal R}$ with boundary $\Sigma = \partial {\cal R}$,
we take the renormalized $S$ to be computed from \(W[g_{ab}]\) via the replica trick, as the response of \(W[g_{ab}]\) to a conical singularity at $\Sigma$ (henceforth called the entangling surface):
\begin{align}\label{Sdef}
	S &= - {\rm Tr} \rho \log \rho = (1 - \partial_{n}) \log {\rm Tr} \rho^{n}.
\end{align}
The density matrix can be written in terms of the path integral, \(\rho^{n} = \frac{Z[g_{(n)ab}]}{Z^{n}[g_{(1)ab}]}\), where \(g_{(n)ab}\) denotes the geometry with \(n\) replicas of the original geometry, glued together at the entangling surface. Thus \(S\) can be expressed in terms of \(W[g_{(n)ab}]\) as
\begin{align}
	S =  W_{(1)} - \partial_{n} W_{(n)}\big|_{n=1},
\end{align}
with \(W_{(n)} \equiv W[g_{(n)ab}]\).

It then remains to compute $S''$.  Given a null congruence $N$ orthogonal to $\Sigma$ (by convention taken outgoing relative to the region ${\cal R}$), we may vary $\Sigma$ by displacing it along the null generators of $N$.  This defines an associated deformation of ${\cal R}$, and thus a change of $S$.  The quantity $S''$ is well-defined at $p$ when the second such variation takes the form
\begin{equation}
\label{eq:Spp}
\frac{\delta }{\delta \Sigma \left(y_1\right)} \left[\frac{1}{\sqrt{h}}  \frac{\delta }{\delta \Sigma \left(y_p\right)}S \right] = \sqrt{h} \ S'' (y_p) \delta(y_1-y_p) + f(y_p,y_1)
\end{equation}
for smooth functions $S''(y_p), f(y_p,y_1)$.  Here $y$ labels the generators of the null congruence $N$, $y_p$ is the null generator through $p$,  and $h$ denotes the determinant of the metric on $\Sigma$ in the $y$-coordinate system.\footnote{In general, one might expect even more singular terms (involving e.g. derivatives of delta-functions) to appear in \eqref{eq:Spp}.  In such cases a QNEC of the form \eqref{QNECintro} cannot hold.}

\subsection{The QNEC and local counter-terms}
\label{subsec:dotheycancel}

The renormalized effective action $W$ depends on the choice of renormalization scheme, though any two schemes will differ only by adding finite local counter-terms; i.e., by the addition to $W$ of integrals of marginal or relevant operators built from background curvatures of $g_{ab}$ and/or matter fields.  Such terms might in principle affect either side of \eqref{QNECintro}.  Below, we calculate the net effect on the QNEC quantity
\begin{equation}
\label{eq:Q}
Q := T_{ab}k^ak^b - \frac{1}{2\pi}S''
\end{equation}
at points $p$ where $k^c$ generates a locally stationary null congruence in a background satisfying the null convergence condition $R_{ab} k^a k^b \ge 0$.  In particular, as noted in the introduction we must have
\begin{equation}
\label{Rkkvanishes}
(R_{ab} k^a k^b)|_p=0
\end{equation}
at all such points.

We consider a theory that approaches a (unitary) conformal fixed point in the UV.  The possible terms thus depend on the spacetime dimension $d$.  We will assume in all cases that there are no scalar operators saturating the unitarity bound $\Delta =\frac{d-2}{2}$, so the addition to $W$ of kinetic terms like $\int d^dx\sqrt{-g} (\partial \phi)^2$ are not allowed.  For simplicity, for $d=2$ we also neglect conserved currents as in this case they would require special treatment.  Apart from this one case, in the absence of non-metric sources combining covariance with unitarity bounds forbids the appearance of terms in $W$ involving CFT operators with spin $j \ge 1$.  For later use in section \ref{sec:killingproof} we note that, using
 an argument like that in footnote \ref{Fufoot} below,
 a result of \cite{Wall:2015raa} thus shows that the QNEC is scheme-independent on any bifurcate Killing horizon.

\subsubsection{$d\le 3$}

For $d\le 3$, the terms one may add to $W$ are only
\begin{align}
\label{eq:dl3terms}
	\int d^dx \sqrt{-g} \phi_1\text{, }\int d^dx \sqrt{-g} \,R \phi_2,
\end{align}
for scalar operators $\phi_1,\phi_2$ of dimensions $\Delta_1,\Delta_2$ with $\Delta_1 \le d$ and $\Delta_2 \le d-2$. Note that this includes the case $\phi_2 = \mathds{1}$. The contribution of the first term to $T_{ab}$ is proportional to $g_{ab}$ and thus vanishes when contracted with $k^a k^b$ for null $k^c$.  Its explicit contribution to $S$ also vanishes, so it does not affect the QNEC.

For terms of the second form in \eqref{eq:dl3terms} one finds
\begin{eqnarray}
\Delta S &=& 4 \pi \int_{C_B} d^{d-2}y \sqrt{h} \phi_2, \\
\Delta T_{ab} &=&  2  \left(\nabla_a \nabla_b - g_{ab} \nabla^2\right) \phi_2  -2  \left({R}_{ab}  - \frac{1}{2}g_{ab} {R}  \right)  \phi_2 .
\end{eqnarray}
We use equation \eqref{eq:Spp} to calculate the derivative of $\Delta S$. The first derivative of $\Delta S$ is $\Delta S'\equiv \frac{1}{\sqrt{h}}\frac{\delta }{\delta \Sigma \left(y_p\right)}\Delta S=4\pi \left(\theta \phi _2+\dot \phi _2\right)$, where the definition of the expansion $\theta :=\frac{1}{\sqrt{h}}\frac{\delta \sqrt{h}}{\delta \Sigma \left(y_p\right)}$ (equation (1.1) of \cite{Fu:2017lps}) is used. The second derivative of the entropy is then $\Delta S''=4\pi \left(\dot \theta \phi _2+\theta \dot \phi _2+\ddot \phi_2\right)$. Since $k^a$ is null, the $kk$-component of $\Delta T_{ab}$ is $\Delta T_{kk}=2\ddot \phi _2-2R_{kk}\phi _2=2\ddot \phi _2+2\phi _2\left(\dot \theta +\frac{\theta ^2}{d-2}+\sigma_{ab}\sigma^{ab}\right)$, where the Raychaudhuri equation \eqref{eq:Ray} is used. Thus, the change of the quantity \eqref{eq:Q} is
\begin{equation}
\Delta Q := \Delta T_{kk} - \frac{1}{2\pi }\Delta S'' = 2 \phi_2 \left( \frac{\theta^2}{d-2} + \sigma_{ab} \sigma^{ab}\right) - 2\theta \dot{\phi_2}.
\end{equation}
Both terms vanish on a locally stationary horizon, and in fact $\sigma_{ab}$ vanishes identically for $d=3$. So under the above conditions the QNEC is scheme-independent for $d \le 3$.  In fact, we see that it is really only necessary to impose $\theta=0$.

\subsubsection{$d=4,5$}
\label{SI45}

Increasing $d$ leads to additional terms.  The allowed terms for $d=4,5$ are those in \eqref{eq:dl3terms} together with
\begin{align}
\label{eq:d45}
\int d^dx \sqrt{-g} \,R_{ab}R^{ab}\text{, }\int d^dx\sqrt{-g} R_{abcd}R^{abcd}.
\end{align}
Since scalar operators $\phi$ have dimensions larger than $\frac{d-2}{2} \ge 1$, for $d=4,5$ they cannot be inserted into the terms \eqref{eq:d45}.   Note that terms like $\int d^dx \sqrt{-g} R^2$ can be written as the second term in \eqref{eq:dl3terms} by taking $\phi_2$ to involve $R$, so such terms were already considered above.

The contributions of \eqref{eq:d45} to the QNEC quantity $Q$ are complicated and do not appear to vanish at the desired points $p$. Indeed, for $d\ge 5$ the results of \cite{Fu:2017lps} show that the particular combination of the terms in \eqref{eq:d45} with $\int d^dx \sqrt{-g} R^2$ that defines the Gauss-Bonnet term contributes $\Delta Q \neq 0$ even on Ricci-flat backgrounds\footnote{\label{Fufoot} Ref. \cite{Fu:2017lps} considered a perturbative computation of $Q_{\text{QFC}} :=\dot{\theta} +4GS''_{\text{GB}}$, where $\dot{\theta}$ is affine derivative of the expansion of $N$.  The computation was done at first order in the Gauss-Bonnet coupling $\gamma$ about a Ricci-flat background.
From the Raychauduri equation \eqref{eq:Ray}, the first order change in $\dot{\theta}$ is precisely $-\Delta T_{ab}k^a k^b$ where $\Delta T_{ab}$ is the Gauss-Bonnet term's contribution to  $T_{ab}$.  Thus $Q_{\text{QFC}} = -4GQ$ with $Q$ defined by \eqref{eq:Q}.}, though the Gauss-Bonnet contribution to $Q$ vanishes for $d=4$.

The fact that terms in \eqref{eq:d45} are four-derivative counter-terms suggests that their contributions to $\Delta T_{ab} k^a k^b$ and $\Delta S''$ contain fourth derivatives of the horizon generator $k^a$.  It is thus natural to ask if we can force $\Delta Q=0$ for \eqref{eq:d45} by setting the first, second, and third derivatives of the expansion $\theta$ and shear $\sigma_{\alpha \beta}$ to zero, where
here and from now on, we use indices $\alpha , \beta , \ldots $ to indicate coordinates on $\Sigma$. It turns out that this is the case. For $d=4,5$ we impose the conditions
\begin{equation}
\label{eq:highD}
\begin{aligned}
\theta |_p&=(\mathscr{D}_a\theta )|_p=(\mathscr{D}_b\mathscr{D}_a\theta )|_p=(\mathscr{D}_c\mathscr{D}_b\mathscr{D}_a\theta )|_p=0,
\\
\sigma _{\alpha \beta }|_p&=(\mathscr{D}_a\sigma _{\alpha \beta })|_p=(\mathscr{D}_b\mathscr{D}_a\sigma _{\alpha \beta })|_p=(\mathscr{D}_c\mathscr{D}_b\mathscr{D}_a\sigma _{\alpha \beta })|_p=0,
\end{aligned}
\end{equation}
where $\mathscr{D}_a$ is the covariant derivative along the congruence $N$. We will show below that (when combined with a positive energy condition) these requirements suffice to show $\Delta Q =0$, though the question remains open which conditions are precisely necessary.

A final condition we impose is that the background solve the Einstein equations with a source respecting the dominant energy condition (DEC) up to a term proportional to the metric.\footnote{This condition is motivated by the discussion of weakly isolated horizons proposed in \cite{Ashtekar:2001jb}.} This is equivalent to requiring $R_{ab}$ to be of the form
\begin{equation}
\label{gDEC}
R_{ab}=R_{ab}^\text{(DEC)}+\alpha g_{ab},
\end{equation}
for some scalar field $\alpha$,
where  for any future-pointing causal (either timelike or null) vector field $v^a$, the vector field $-R^a_{~b}{}^\text{(DEC)}v^b$ must also be both future-pointing and causal.  A short argument (see appendix \ref{appendix:identities}) using \eqref{eq:highD} then shows that on the null generator through $p$ we have
\begin{align}\label{Rak}
	R_{ab}k^{b} = fk_{a} + O(\lambda^3),
\end{align}
for some scalar function $f$ and that
\begin{align}
\label{Rkk}
	R_{abcd}k^{b} k^{d} = \zeta k_{a} k_{c} + O(\lambda^3),
\end{align}
for some scalar function \(\zeta \). Since \eqref{eq:highD} implies that equation \eqref{nablak} holds at point $p$, the contracted Bianchi identity implies that equation \eqref{eq:WIHBianchi} holds at point $p$, namely
\begin{align}
\label{eq:Bianchi}
	(k^a \partial_{a} f)|_p  = (\frac{1}{2} k^a \partial_{a}R)|_p = (- k^a \partial_{a} \zeta)|_p.
\end{align}

The expressions for $\Delta S$ for the counter-terms \eqref{eq:d45} can be found in \cite{Dong:2013qoa} and the expressions for $\Delta T_{ab}$ for these counter-terms can be calculated by using the definition \eqref{Tdef}. These expressions are simplified greatly by using conditions \eqref{eq:highD} and \eqref{gDEC}. After such simplifications it is straightforward (see appendix \ref{appendix:schemeindependence}) to compute $\Delta S''$ and, as shown in table \ref{tb:4dcounterterm}, the above conditions suffice to force $\Delta Q=0$ for both terms \eqref{eq:d45} in all dimensions. In table \ref{tb:4dcounterterm} we have used the notation $\partial_k := k^a \partial_a$, which we continue to use elsewhere below.

As a final comment, we note that a careful analysis of the calculation shows that although for $d=5$ we require the full list of conditions \eqref{eq:highD} show $\Delta Q=0$, for $d=4$ it suffices to use only a subset of the conditions. The reason is that for $d=4$ we may choose to study the Gauss-Bonnet term $R_{abcd} R^{abcd} - 4 R_{ab}R^{ab} + R^2$ (instead of $R_{abcd} R^{abcd}$).  This term is topological and so contributes to neither $T_{ab}$ nor $S''$, and to guarantee $\Delta Q=0$ for the only remaining counter-term $R_{ab}R^{ab}$ we need only \eqref{gDEC} and conditions on the first line of \eqref{eq:highD}.

\begin{table}[]
\centering
\renewcommand{\arraystretch}{1.5}
\caption{Scheme-independence of QNEC for all four-derivative counter-terms $\Delta L$ from \eqref{eq:d45} when \eqref{eq:highD} and \eqref{gDEC} hold.  \protect\\}
\label{tb:4dcounterterm}
\begin{tabular}{|c|c|c|c|}
\hline
$\Delta L$ & $\Delta T_{ab}k^ak^b$ & $\Delta S$ & $\Delta Q$ \\ \hline
$R^2$ & $4\partial _k^2R$ & $8\pi R$ & $0$ \\ \hline
$R^{ab}R_{ab}$ & $4\partial _k^2f$ & $8\pi f$ & $0$ \\ \hline
$R^{abcd}R_{abcd}$ & $-8\partial _k^2\zeta $ & $-16\pi \zeta $ & $0$ \\ \hline
\end{tabular}
\end{table}

\subsubsection{$d \ge 6$}\label{scheme-indep}

In six dimensions, six-derivative counter-terms become allowed. Based on the results above, one might hope to maintain scheme-independence of the QNEC in this case by requiring even more derivatives of the extrinsic curvature to vanish. However, we now show that for $d\ge 6$ the contribution $\Delta Q$ is generally non-zero even on weakly isolated horizons (where $\theta, \sigma_{\alpha \beta}$ vanish identically on $N$  \cite{Ashtekar:2001jb})  in backgrounds satisfying \eqref{gDEC}.

Since all derivatives of $\theta, \sigma_{\alpha\beta}$ vanish, the results \eqref{Rak}, \eqref{Rkk}, and \eqref{eq:Bianchi} now exactly hold on a finite neighborhood of point $p$ on the horizon. Furthermore, we show in appendix \ref{appendix:identities} that on weakly isolated horizons the Riemann tensor $R_{abck}$ can be written as $R_{abck}=k_c\tilde{A}_{ab}+k_{[a}\tilde{B}_{b]c}$, where $\tilde{A}_{ab}$ is antisymmetric and satisfies $k^a\tilde{A}_{ab}\propto k_b$ and $\tilde{B}_{ab}$ satisfies $k^a\tilde{B}_{ab}\propto k_b$ and $k^b\tilde{B}_{ab}\propto k_a$. This allows one to write down additional relations also listed in appendix \ref{appendix:identities}.  Together, they allow one to show the QNEC to be unchanged by adding six-derivative counter-terms built from polynomial contractions of the Riemann tensor.  The computations are presented in appendix \ref{appendix:schemeindependence} and the results are summarized in table \ref{tb:6dcounterterm}.

\begin{table}[]
\centering
\renewcommand{\arraystretch}{1.5}
\caption{Scheme-independence of QNEC for the six-derivative counter-terms $\Delta L$ built from polynomial contractions of the Riemann tensor when the null congruence $N$ is a weakly isolated horizon. However, as shown in the main text, scheme-independence can fail for counter-terms involving derivatives of the Riemann tensor.  \protect\\}
\label{tb:6dcounterterm}
\begin{tabular}{|c|c|c|c|}
\hline
$\Delta L$ & $\Delta T_{ab}k^a k^b$ & $\Delta S$ & $\Delta Q$ \\ \hline
$R^3$ & $6\partial _k^2(R^2)$ & $12\pi R^2$ & $0$ \\ \hline
$RR^{ab}R_{ab}$ & $2\partial _k^2(R^{ab}R_{ab})+4\partial _k^2(Rf)$ & $4\pi R^{ab}R_{ab}+8\pi Rf$ & $0$ \\ \hline
$R^{ab}R_{bc}R^c_{~a}$ & $6\partial _k^2(f^2)$ & $12\pi f^2$ & $0$ \\ \hline
$R^{abcd}R_{ac}R_{bd}$ & $2\partial _k^2(f^2)-4\partial _k^2(R_{akcl}R^{ac})$ & $4\pi f^2-8\pi R_{akcl}R^{ac}$ & $0$ \\ \hline
$RR^{abcd}R_{abcd}$ & $2\partial _k^2(R^{abcd}R_{abcd})-8\partial _k^2(R\zeta )$ & $4\pi R^{abcd}R_{abcd}-16\pi R\zeta $ & $0$ \\ \hline
$R^{abcd}R_{cdbe}R^e_{~a}$ & $-8\partial _k^2(\zeta f)-2\partial _k^2(R_k^{~bcd}R_{cdbl})$ & $-16\pi \zeta f-4\pi R_k^{~bcd}R_{cdbl}$ & $0$ \\ \hline
$R^{ab}_{~~cd}R^{ce}_{~~bf}R^{df}_{~~ae}$ & $-6\partial _k^2(R_{kel}^{~~~f}R_{lfk}^{~~~e})+6\partial _k^2(\zeta ^2)$ & $-12\pi R_{kel}^{~~~f}R_{lfk}^{~~~e}+12\pi \zeta ^2$ & $0$ \\ \hline
$R^{abcd}R_{cdef}R^{ef}_{~~ab}$ & $-12\partial _k^2(R^{ef}_{~~lk}R_{eflk})$ & $-24\pi R^{ef}_{~~lk}R_{eflk}$ & $0$ \\ \hline
\end{tabular}
\end{table}

However, we will shortly see that scheme-independence of the QNEC can fail even on weakly isolated horizons for counter-tems that contain derivatives of the Riemann tensor. There are four such counter-terms, $(\nabla _aR)(\nabla ^aR)$, $(\nabla _aR_{bc})(\nabla ^aR^{bc})$, $(\nabla _eR_{abcd})(\nabla ^eR^{abcd})$, and $(\nabla _aR_{bc})(\nabla ^bR^{ac})$. Neglecting total derivatives of the action, these counter-terms are not linearly independent and one can write the last two previous counter-terms in terms of the other ten \cite{Oliva:2010zd}.

We will show this failure for the term $(\nabla _aR)(\nabla ^aR)$. From \cite{Oliva:2010zd} we have
\begin{equation}
\begin{aligned}
\Delta T_{kk}&= k^ak^b \Delta T_{ab}\\
&= -4k^ak^b\nabla _a\nabla _b(\nabla _e\nabla ^eR)-2(k^a\nabla _aR)(k^b\nabla _bR)\\
&=-4\partial _k^2(\nabla _a\nabla ^aR)-2(\partial _kR)^2.
\end{aligned}
\end{equation}
The formula used to calculate the change in the entropy is given in \cite{Miao:2014nxa}, where it was written as
\begin{equation}
\label{eq:DeltaSfor6d}
	\Delta S = \Delta S_{\text{G-Wald}} + \Delta S_{\text{Anomaly}}.
\end{equation}
On weakly isolated horizons, the ``generalized Wald entropy'' term in \eqref{eq:DeltaSfor6d} reduces to the ordinary Wald entropy and is given by
\begin{equation}
\Delta S_{\text{G-Wald}}=\Delta S_\text{Wald}=-2\pi \times (-2\nabla _a\nabla ^aR)\times (-2)=-8\pi \nabla _a\nabla ^aR.
\end{equation}
On weakly isolated horizons, by applying equation (A.12) of \cite{Miao:2014nxa}, one finds that the ``entropy anomaly'' term in \eqref{eq:DeltaSfor6d} vanishes for counter-term $(\nabla _aR)(\nabla ^aR)$. Thus,
\begin{equation}
\Delta S_{\text{Anomaly}}=0.
\end{equation}
Finally, we obtain
\begin{equation}
\label{eq:DQ6d}
\Delta Q=\Delta T_{kk}-\frac{1}{2\pi }\Delta S''=-2(\partial _kR)^2.
\end{equation}
This quantity does not vanish generally, although it vanishes on Ricci-flat background.

It is not hard to find an explicit geometry in which this counter-term spoils the scheme-independence of the QNEC. Consider the spacetime metric
\begin{equation}
ds^2=-dudv-dvdu-cu^2v^2du^2+\sum\limits_{\alpha }{{{\left( d{{y}_{\alpha }} \right)}^{2}}},
\end{equation}
where $c$ is a positive constant which is not assumed to be small. In this spacetime, there is a non-expanding null surface $v=0$.
Its Ricci tensor is
\begin{equation}
\label{eq:exRab}
R_{ab}=c^2u^4v^2(du)_a(du)_b+cu^2(du)_a(dv)_b+cu^2(dv)_a(du)_b = -cu^2 g_{\perp ab},
\end{equation}
where $g_{\perp ~b}^{~a}$ is the projector onto the $u$-$v$ plane. Since this plane is timelike, for any future-pointing causal vector $v^b$ the vector $-R^a_{~b} v^b = cu^2 g^{~a}_{\perp ~b} v^b$ is again future-pointing and causal and the Ricci tensor \eqref{eq:exRab} satisfies \eqref{gDEC} with $\alpha =0$. Thus the plane $v=0$ is a weakly isolated horizon.  But from \eqref{eq:exRab} we find scalar curvature $R=-2cu^2$, hence
\begin{equation}
\Delta Q=-32c^2u^2\neq 0
\end{equation}
and the QNEC fails to be scheme-independent.

\section{Holographic Proofs of the QNEC}
\label{sec:HologProofs}

The QNEC was proven to hold in \cite{Koeller:2015qmn} for leading-order holographic field theories on flat spacetimes.   We review this derivation in section \ref{subsec:outline} below and show that the argument admits a straightforward generalization to arbitrary curved backgrounds; i.e., to the case where the boundary of the asymptotically locally AdS bulk is arbitrary.  However, the resulting inequality is generally divergent, and we expect it to yield a finite renormalized QNEC only in the contexts where the QNEC is scheme-independent.  For the scheme-independent cases described in section \ref{sec:schemedep}, we will indeed be able to derive such a finite renormalized QNEC below.

Since all the proofs in this section are provided in the context of AdS/CFT correspondence, we change our index notation to make it more suitable for this context. In this section, we use indices $\mu ,\nu ,\ldots $ to indicate the $d+1$ coordinates on the bulk spacetime and use indices $i,j,\ldots $ to indicate the $d$ coordinates on the boundary spacetime.

\subsection{Outline of holographic proofs}
\label{subsec:outline}

The central idea of \cite{Koeller:2015qmn} is to reformulate both \(T_{kk}\) and \(S''\) in terms of quantities in the dual bulk asymptotically AdS spacetime, and to use a fact about extremal surfaces known as ``entanglement wedge nesting'' (EWN) \cite{Czech:2012bh,Wall:2012uf,Akers:2016ugt} to provide the desired inequality. To begin, consider two regions \(A,B\) on the asymptotically AdS boundary. Entanglement wedge nesting states that if these boundary regions are nested in the sense that \(D(B) \subseteq D(A)\) then  their extremal surfaces \(e(A)\), \(e(B)\) must also be nested, i.e. everywhere spacelike related. Here
 \(D(A)\) is the domain of dependence of \(A\) in the asymptotically AdS boundary.

Now consider a family of boundary regions \(A(\lambda)\) with entangling surfaces \(\partial A(\lambda)\), which differ by localized deformations along a single null generator \(k^{i}(y)\) of a null hypersurface shot out from the initial surface \(A(0)\).  The derivatives of the entropy used in the QNEC are then derivatives with respect to this particular \(\lambda\), which we can take to be an affine parameter for \(k^{i}\). Consider the codimension-1 bulk surface \(\mathcal{M}\) foliated by the (smallest for each $A(\lambda)$) extremal surfaces \(e(\lambda) \equiv e(A(\lambda))\),
\begin{align}
	\mathcal{M}:= \cup_{\lambda} e(\lambda)
\end{align}
EWN for \(A(\lambda)\) implies that \(\mathcal{M}\) is a spacelike surface.

The surface \(\mathcal{M}\) can be parametrized by \(\lambda\) and the \(d-1\) coordinates \(y^{\alpha} = \{z,y^{a}\}\) associated with each \(e(\lambda)\). The coordinate basis for the tangent space of \(\mathcal{M}\) then consists of \(\partial_{\lambda}X^{\mu}\) and the \(d-1\) coordinate basis tangent vectors of the extremal surfaces, \(\partial_{\alpha} X^{\mu}\). By EWN, each of these vectors has positive norm. The norm of \(\partial_{\lambda}X^{\mu}\) can be expanded in \(z\) near the boundary, and will involve the near-boundary expansion of both the metric and the extremal surface embedding functions. EWN implies in particular that as \(z\to 0\) the most dominant term in this expansion is positive. In \cite{Koeller:2015qmn} it was shown that for locally-stationary surfaces (satisfying \(\theta |_p= \sigma _{ab}|_p= 0\) at a point $p$) in flat space one has\footnote{Any vector tangent to \(\mathcal{M}\) has positive norm. The original proof of \cite{Koeller:2015qmn} used \(s^{\mu} \equiv t^{\mu}_{~\nu} \partial_{\lambda}X^{\nu}\), where \(t^{\mu}_{~\nu}\) projects onto the 2-dimensional subspace orthogonal to \(e(\lambda)\), instead of \(\partial_{\lambda}X^{\mu}\) itself. Both work equally well to derive the QNEC in \(d\geq3\). We restrict to \(d\geq3\) and use \(\partial_{\lambda}X^{\mu}\) for simplicity. Note that for $d=2$ by a change of conformal frame one may always choose to work on a flat background.}
\begin{align}\label{flatchi^{2}}
	0 \leq G_{\mu\nu} \partial_{\lambda}X^{\mu} \partial_{\lambda}X^{\nu} = \frac{16\pi G}{d} z^{d-2} \left( T_{kk} - \frac{1}{2\pi} S'' \right),
\end{align}
where the quantities on the right-hand side are have been renormalized.\footnote{In flat space, the local stationarity condition makes the renormalization trivial; \(T_{kk}\) and \(S''\) are finite to begin with \cite{Koeller:2015qmn}. This is not guaranteed if the boundary spacetime is curved.} Thus for these surfaces in flat space, EWN implies the renormalized QNEC.

As indicated in equation \eqref{flatchi^{2}}, the renormalized quantities in the QNEC appear in the EWN inequality at \(O(z^{d-2})\). So in order to derive the renormalized QNEC, the terms in the \(z\)-expansion of \((\partial_{\lambda}X^{\mu})^{2}\) must vanish at all lower orders. This condition provides restrictions on the surfaces and space-times. In flat space, it is sufficient to have local stationarity, i.e. $\theta |_p=0, \sigma_{\alpha \beta}|_p=0$ at a point $p$ \cite{Koeller:2015qmn}. More generally the condition may be more complicated. To compute \eqref{flatchi^{2}} explicitly, we set $\ell_{\text{AdS}}=1$ and use Fefferman-Graham-style coordinates to introduce the near-boundary expansion of the bulk metric $G_{\mu \nu}$:
\begin{equation}
\begin{aligned}
	G_{zz}(x^{\mu}) &= \frac{1}{z^{2}}\text{, }G_{zi} = 0,\\
	\label{Gexpn}
	G_{ij}(x^{\mu}) &\equiv \frac{1}{z^{2}} g_{ij} = \frac{1}{z^{2}} \left( g_{(0)ij} + g_{(2)ij} + \ldots + g_{(dl)ij} + \bar{g}_{(d)ij} + \frac{16\pi G}{d} z^{d}T_{ij} \right),
\end{aligned}
\end{equation}
and the embedding functions of the extremal surfaces
\begin{equation}
\label{Xexpn}
\begin{aligned}
	X^{z}(y^{\alpha}) &= z, \\
	X^{i}(y^{\alpha}) &= X^{i}_{(0)} + X^{i}_{(2)} + \ldots + X^{i}_{(dl)} -\frac{4G}{d} z^{d} g_{(0)}^{ij} S'_{i} + \bar{X}^{i}_{(d)},
\end{aligned}
\end{equation}
where
\begin{align}\label{S'idef}
	S'_{i} \equiv \frac{1}{\sqrt{h_{(0)}}} \frac{\delta S_{\rm ren}}{\delta X^{i}_{(0)}}
\end{align}
is the renormalized entropy directional derivative per unit area and
subscripts denote powers of \(z\), e.g. \(X^{i}_{(m)}\) is \(O(z^{m})\), while \(g_{(dl)ij}\) is a log term of order \(z^{d}\log{z}\). As a result, $\partial _\lambda X^\mu$ involves $S'' = \partial_\lambda(k^i S_i{}')$. The terms \(\bar{g}_{(d)ij}\) and \(\bar{X}^{i}_{(d)}\) refer to the geometric parts of the \(O(z^{d-2})\) and \(O(z^{d})\) parts of the metric and embedding functions, respectively.
For more details, see \cite{Hung:2011ta,Koeller:2015qmn,Schwimmer:2008yh} for the extremal surface expansion, and e.g. \cite{deHaro:2000vlm,Hung:2011ta} for the metric.

The key point is that \(S' \equiv k_{(0)}^{i}S'_{i}\) and \(T_{ij}k^{i}_{(0)}k^{i}_{(0)}\) appear at \(O(z^{d-2})\). This is why plugging these expansions into \((\partial_{\lambda}X^{\mu})^{2}\) gives the QNEC at \(O(z^{d-2})\), as in \eqref{flatchi^{2}}. That the CFT stress tensor appears at \(O(z^{d-2})\) in the near-boundary expansion for the metric is well known. We now derive the appearance of \(S'_{i}\) in equation \eqref{Xexpn}.

In Einstein-Hilbert gravity, the HRT prescription \cite{Hubeny:2007xt} is
\begin{align}
\label{eq:HRTaction}
	S(\lambda) = \frac{A(e(\lambda))}{4G} =  \frac{1}{4G} \int dz d^{d-2}y \sqrt{H[X]},
\end{align}
where $H[X]$ is the determinant of the induced metric on $e(\lambda)$ written as a functional of $X$. Varying the on-shell area functional with respect to \(X^{i}\) gives a boundary term evaluated at a cutoff surface \(z = \rm const\). This produces the regulated entropy variation
\begin{align}\label{deltaS}
	\frac{1}{\sqrt{h}}\frac{\delta S_{\rm reg}}{\delta X^{i}} = - \frac{1}{4G} z^{1-d} \frac{g_{ij} \partial_{z} X^{j}}{\sqrt{1+g_{nm}\partial_{z}X^{n}\partial_{z}X^{m}}} \Bigg|_{z=\rm const},
\end{align}
where $h$ is the determinant of the induced metric. Everything in equation \eqref{deltaS} (including \(\delta X^{i}\)) is to be expanded in \(z\) and evaluated at a cutoff surface at \(z=\rm const\). In general, there will be terms that diverge as \(z \to 0\).

The entropy can be renormalized using in a manner similar to that used for the on-shell action (see e.g. \cite{deHaro:2000vlm}). Indeed, the two are intimately related \cite{Taylor:2016aoi}. Adding local, geometrical, covariant counter-terms gives the renormalized entropy via
\begin{align}
	S_{\rm ren} = \lim_{z\to 0} \,\left(S_{\rm reg} + S_{\rm ct}\right).
\end{align}
With an arbitrary choice of renormalization scheme, \(S_{\rm ct}\) can contain finite counter-terms in addition to those required to cancel divergences. Expanding \eqref{deltaS} and removing the divergences, we have the finite renormalized entropy variation given in general by
\begin{align}\label{Srenexpn}
	\frac{1}{\sqrt{h_{(0)}}}\frac{\delta S_{\rm ren}}{\delta X^{i}_{(0)}} = -\frac{d}{4G z^{d}} g_{(0)ij} X^{j}_{(d)} + \ldots
\end{align}
where $h_{(0)}$ is the determinant of the metric induced on $\Sigma$ by $g_{(0)ij}$, \(X^{j}_{(d)}\) denotes the \(O(z^{d})\) part of \(X^{j}\), and the ``\(\ldots\)'' denotes finite, local, geometric terms, which include both finite contributions to equation \eqref{deltaS} from products of lower-order terms in the embedding functions, as well as possible finite counter-terms from \(S_{\rm ct}\). Re-arranging for \(X^{i}_{(d)}\) gives
\begin{align}\label{X^{i}_{d}}
	X^{i}_{(d)} = -\frac{4G}{d} z^{d} g_{(0)}^{ij} \frac{1}{\sqrt{h_{(0)}}} \frac{\delta S_{\rm ren}}{\delta X^{j}_{(0)}} + \bar{X}^{i}_{(d)}
\end{align}
In this expression, \(\bar{X}^{i}_{(d)}\) contains the contribution from the ``\(\ldots\)'' of equation \eqref{Srenexpn}. Plugging \eqref{X^{i}_{d}} into the expansion of \(X^{i}\) yields equation \eqref{Xexpn}, as promised.

The geometric restrictions on the surface and geometry which guarantee that the lower-order terms in expression \eqref{flatchi^{2}} vanish can in principle be determined by solving the relevant equations, though we will not carry out an exhaustive analysis here.

We show in section \ref{subsec:Hd=3} below that the above argument leads to a finite renormalized scheme-independent QNEC for $d\le 3$ at points $p$ where the expansion $\theta$ vanishes for the chosen null congruence $N$. We then show in \ref{subsec:Hd=4} and \ref{subsec:Hd=5} below that for $d=4,5$ it leads to a finite renormalized scheme-independent QNEC at points $p$ where the chosen null congruence $N$ satisfies the conditions on the first line of \eqref{eq:highD} on backgrounds satisfying \eqref{gDEC}. From the results of section \ref{SI45} and the fact \cite{deHaro:2000vlm} that -- for Einstein-Hilbert gravity in the bulk -- holographic renormalization requires only counter-terms that can be built from the Ricci tensor, it is no surprise that we do not require the full list of conditions \eqref{eq:highD}. Finally, in section \ref{sec:killingproof} we provide a finite renormalized scheme-independent QNEC for holographic theories on Killing horizons in arbitrary backgrounds.

\subsection{Proof of the $d\le3$ holographic QNEC}
\label{subsec:Hd=3}
For $d=3$
the asymptotic metric expansion \eqref{Gexpn} and the asymptotic embedding function expansion \eqref{Xexpn} take the form
\begin{align}\label{eq:expns3d}
	g_{ij}(x,z) &= g_{(0)ij} + g_{(2)ij} + g_{(3)ij} + \ldots , \\
	X^i(x,z) &= X^i_{(0)} + X^i_{(2)} + X^i_{(3)} +\ldots .
\end{align}
The causal property of extremal surfaces then implies
\begin{equation}
\begin{aligned}
\label{eq:inequality3}
	0 &\leq g_{ij}(\partial_{\lambda}X ^i)(\partial_{\lambda}X^j) \\
	&= g_{(0)ij}k^{i}k^{j} + g_{(2)ij}k^{i}k^{j} + 2 g_{(0)ij} (\nabla_{\lambda} X_{(2)}^{i}) k^{j} + g_{(3)ij} k^ik^j + 2 g_{(0)ij} (\nabla_{\lambda}X^i_{(3)})k^j.
\end{aligned}
\end{equation}
One can easily check that the Einstein equations and extremal surface equation at second order in \(z\) give
\begin{align}\label{g2X2}
	g_{(2)ij} = \frac{z^{2}}{d-2} \left( R_{ij} - \frac{1}{2(d-1)} R \, g_{(0)ij}\right)~, \quad X_{2}^{i} = \frac{1}{2(d-2)} z^{2} K^{i}~,
\end{align}
in terms of the (traced) extrinsic curvature $K^i : = g_{(0)}^{jk} K^i_{~jk}$ of the (boundary) codimension-2 surface $\partial A$ with conventions given by \eqref{eq:DefExtrCurv}. Since $k^i$ is null, $ \dot{\theta} = \nabla_{\lambda}(g_{(0)ij} K^{i}k^{j})$ and that $k^i$ satisfies the geodesic equation $\nabla_\lambda k^i := k^j \nabla_j k^i = 0$, the terms on the second line of \eqref{eq:inequality3} combine to give
\begin{align}
\frac{z^2}{d-2} \left(  R_{ij} k^i k^j 	+ \dot{\theta} \right) = - \frac{z^2}{d-2} \left(  \frac{\theta^2}{d-2} + \sigma^{ij}\sigma_{ij}\right),
\end{align}
where in the last step we have used \eqref{eq:Ray}.  Both terms vanish on a locally stationary horizon, and in fact $\sigma_{ij}$ vanishes identically for $d=3$.  The terms on the third line of \eqref{eq:inequality3} then give the renormalized QNEC \eqref{flatchi^{2}}.

The $d=2$ argument is identical in form.  Though terms at order $z^2$ are not divergent for $z=2$, there are then divergent terms at order $z^2 \log z$ which are structurally the same as the $z^2$ terms for $d=3$.

\subsection{Proof of the $d=4$ holographic QNEC}

\label{subsec:Hd=4}

For the case of four dimensional boundary, the asymptotic metric expansion \eqref{Gexpn} and the asymptotic embedding function expansion \eqref{Xexpn} take the form
\begin{align}\label{eq:expns}
	g_{ij}(x,z) &= g_{(0)ij} + g_{(2)ij} + g_{(4l)ij} + g_{(4)ij} + \ldots , \\
	X^i(x,z) &= X^i_{(0)} + X^i_{(2)} + X^i_{(4l)} + X^i_{(4)} +\ldots .
\end{align}
The causal property of extremal surfaces provides us the inequality in QNEC. We have
\begin{equation}
\begin{aligned}
\label{eq:inequality}
	0 &\leq g_{ij}(\partial_{\lambda}X ^i)(\partial_{\lambda}X^j) \\
	&= g_{(0)ij}k^{i}k^{j} + g_{(2)ij}k^{i}k^{j} + 2 g_{(0)ij} (\nabla_{\lambda} X_{(2)}^{i}) k^{j} \\
	&+ g_{(4l)ij} k^ik^j + 2 g_{(0)ij} (\nabla_{\lambda}X^i_{(4l)})k^j + g_{(4)ij} k^ik^j + 2 g_{(0)ij} (\nabla_{\lambda}X^i_{(4)})k^j.
\end{aligned}
\end{equation}
Terms on the second line of \eqref{eq:inequality} vanish just as in section \ref{subsec:Hd=3}. From \cite{deHaro:2000vlm,Fischetti:2012rd} we have
\begin{equation}
\label{eq:g4lkk}
g_{(4l)ij}k^ik^j=-\frac{z^4\log z}{24}\partial _k^2R,
\end{equation}
\begin{equation}
\label{eq:g4kk}
	g_{(4)ij}k^ik^j = z^{4} \left(4\pi GT_{kk}+\frac{1}{32}\partial _k^2R\right),
\end{equation}
which may be used to calculate the third line of equation \eqref{eq:inequality}. Here and below we freely use equation \eqref{nablak} which follows from the first line of \eqref{eq:highD}. We will first show that the log terms cancel each other, and then show that the \(O(z^{4})\) terms produce the QNEC.

We introduce the standard notation
\begin{equation}
\label{Aren}
A_\text{ren} = A_\text{reg} + A_\text{ct}.
\end{equation}
The entropy counter-terms $A_\text{ct}$ generically contain a finite part which we must extract. This comes from the requirement that the counter-terms are covariant functionals of the geometric quantities on the cutoff surface. The counter-term $A_{\text{ct, }O\left(\log z\right)}$ which cancels the log divergence has an explicit $\log z$, and consequently has no finite part. The finite part $A_\text{ct, finite}$ of the counter-terms comes from the counter-term which cancels the leading area-law divergence. This is \cite{Taylor:2016aoi}
\begin{align}
A_\text{ct, A} &= -\frac{1}{d-2} \int_{z=\epsilon} d^{d-2}y\sqrt{\gamma }
= -\frac{1}{2} \int_{z=\epsilon} d^{d-2}y\sqrt{\gamma ^{(0)}} \frac{1}{z^2}\left[ 1 + \frac{1}{2} g^{||ij}g_{(2)ij} + \ldots \right],
\end{align}
where $\sqrt{\gamma }$ is the induced metric on the intersection of the HRT surface with the cutoff surface, and $g^{\parallel ij}$ is the part of the boundary metric parallel to the entangling surface. Using the first equality of \eqref{eq:Bianchi}, the finite part of the counter-term can be written
\begin{equation}
\begin{aligned}
A_\text{ct, finite}&=-\frac{1}{4}\int d^{d-2}y\sqrt{\gamma ^{(0)}}\left(g^{ij}-g^{\perp ij}\right)\left(-\frac{1}{2}R_{ij}+\frac{1}{12}Rg_{ij}\right)\\
&=  -\frac{1}{4}\int d^{d-2}y\sqrt{\gamma ^{(0)}}\left(f-\frac{R}{3}\right).
\end{aligned}
\end{equation}
Thus we find
\begin{equation}
A'_\text{ct, finite}=-\frac{1}{24}\partial _kR.
\end{equation}

From Eq.~\eqref{deltaS} we have
\begin{equation}
\label{A'reg}
A'_\text{reg} = -\frac{1}{z^{d-1}} g_{ij} (\partial _{z}X^i) (\partial _{\lambda }X^j).
\end{equation}
Expanding the above equation and using equation \eqref{g2X2}, we find that the non-trivial contributions are at the same order as in \eqref{eq:inequality}, which are
\begin{equation}
A'_\text{reg} = -\frac{1}{z^{3}} \left[ g_{(0)ij} (\nabla_{z} X^i_{(4l)}) k^j|_{z^3\log z}+g_{(0)ij} (\nabla_{z} X^i_{(4l)}) k^j|_{z^3} + g_{(0)ij} (\nabla_{z} X^i_{(4)}) k^j \right].
\end{equation}
The first term on right-hand side is the logarithmic divergence of the first derivative of the entropy, which can be computed as \cite{Myers:2013lva}
\begin{equation}
A'_{\text{reg, }O\left(\log z\right)}=-\frac{1}{2}(\log z)\partial _k\left(f-\frac{R}{3}\right)=-\frac{1}{12}(\log z)\partial _kR
\end{equation}
From this, we can infer that
\begin{align}
g_{(0)ij}(\nabla _zX^i_{(4l)})k^j|_{z^3\log z}=&\frac{1}{12}z^3(\log z)\partial _kR, \\
g_{(0)ij}X^i_{(4l)}k^j=&\frac{1}{48}z^4(\log z)\partial _kR, \\
g_{(0)ij}(\nabla _zX^i_{(4l)})k^j|_{z^3}=&\frac{1}{48}z^3\partial _kR,\\
g_{(0)ij}(\nabla _{\lambda }X^i_{(4l)})k^j=&\frac{1}{48}z^4(\log z)\partial _k^2R.
\end{align}
As expected, the $O\left(z^4\log z\right)$ terms in equation (\ref{eq:inequality}) cancel:
\begin{equation}
g_{(4l)ij} k^ik^j + 2 g_{(0)ij} (\nabla_{\lambda}X^i_{(4l)})k^j=0.
\end{equation}
And since the rate of change of the renormalized area is
\begin{align}
A'_\text{ren}= A'_\text{reg, finite}+A'_\text{ct, finite}
=  -\frac{1}{z^3}\left[\frac{1}{48}z^3\partial _kR+g_{(0)ij}(\partial _zX^i_{(4)})k^j\right]-\frac{1}{24}\partial _kR,
\end{align}
we obtain
\begin{align}
g_{(0)ij}(\nabla _zX^i_{(4)})k^j= & z^3\left(-A'_\text{ren}-\frac{1}{16}\partial _kR\right),\\
g_{(0)ij}X^i_{(4)}k^j= & \frac{z^4}{4}\left(-A'_\text{ren}-\frac{1}{16}\partial _kR\right),\\
g_{(0)ij}(\nabla _\lambda X^i_{(4)})k^j= & z^4\left(-\frac{A''_\text{ren}}{4}-\frac{1}{64}\partial _k^2R\right). \label{g0X4k}
\end{align}
We now have all we need to evaluate the rest of the final line of equation \eqref{eq:inequality} and derive the QNEC. Plugging \eqref{g0X4k} and equation (\ref{eq:g4kk}) into equation (\ref{eq:inequality}), we obtain the QNEC
\begin{equation}
0\leq T_{kk}-\frac{A''_\text{ren}}{8\pi G}.
\end{equation}

\subsection{Proof of the $d=5$ holographic QNEC}

\label{subsec:Hd=5}
The $d=5$ case is similar.  We find
\begin{equation}
\begin{aligned}
\label{eq:inequality5d}
	0 &\leq g_{ij}(\partial_{\lambda}X^i) (\partial_{\lambda}X^i) \\
	&= g_{(0)ij}k^{i}k^{j} + g_{(2)ij}k^{i}k^{j} + 2 g_{(0)ij} (\nabla_{\lambda} X_{(2)}^{i}) k^{j} \\
	&+ g_{(4)ij} k^ik^j + 2 g_{(0)ij} (\nabla_{\lambda}X^i_{(4)})k^j + g_{(5)ij} k^ik^j + 2 g_{(0)ij} (\nabla_{\lambda}X^i_{(5)})k^j,
\end{aligned}
\end{equation}
with terms on the second line vanishing just as in section \ref{subsec:Hd=3}. As before, we will freely use equation \eqref{nablak} which follows from \eqref{eq:highD}. For $d=5$ we find \cite{deHaro:2000vlm,Fischetti:2012rd}
\begin{equation}
\label{eq:g4kk5d}
g_{(4)ij}k^ik^j=\frac{z^4}{32}\partial _k^2R,
\end{equation}
\begin{equation}
\label{eq:g5kk}
g_{(5)ij}k^ik^j=z^5\frac{16\pi G}{5}T_{kk}.
\end{equation}

We again consider \eqref{Aren}.
Because the boundary dimension is odd, the counter-term
\begin{align}
	A_\text{ct, A} &= -\frac{1}{d-2} \int_{z=\epsilon} d^{d-2}y\sqrt{\gamma }
	= -\frac{1}{3} \int_{z=\epsilon} d^{d-2}y\sqrt{\gamma ^{(0)}} \frac{1}{z^3}\left[ 1 + \frac{1}{2} g^{||ij}g_{(2)ij} + \ldots \right].
\end{align}
contributes no finite part to the renormalized entropy;
\begin{align}
	A_\text{ct, finite} =0.
\end{align}
From Eq.~\eqref{deltaS} we have
\begin{align}\label{Aprime}
	A'_\text{reg} = -\frac{1}{z^{d-1}} g_{ij} (\partial_{z}X^i) (\partial_{\lambda}X^j).
\end{align}
Expanding the above equation, we find that the non-trivial contributions are at the same order as above, which are
\begin{align}
	A'_{\rm reg} = -\frac{1}{z^{4}} g_{(0)ij} (\nabla_{z} X^i_{(4)}) k^j -\frac{1}{z^{4}} g_{(0)ij} (\nabla_{z} X^i_{(5)}) k^j.
\end{align}
The first term on right-hand side is a divergence which must be canceled by the counter-term and the second term on right-hand side is the finite part. We introduce the following notations:
\begin{align}
	A'_{\text{reg, }O(z^{-1})} = -\frac{1}{z^{4}} g_{(0)ij} (\nabla_{z} X^i_{(4)}) k^j, \\
	A'_\text{reg, finite} = -\frac{1}{z^{4}} g_{(0)ij} (\nabla_{z} X^i_{(5)}) k^j.	
\end{align}
Using \eqref{Rak}, the $O(z^{-1})$ divergent part of the regulated area can be computed as \cite{Myers:2013lva}
\begin{equation}
\begin{aligned}
A_{\text{reg, }O(z^{-1})}&=\int d^{d-2}y\sqrt{\gamma ^{(0)}}\frac{1}{2z}\left(\frac{1}{3}R_{ij}g^{\perp ij}-\frac{5}{24}R\right)\\
&=\int d^{d-2}y\sqrt{\gamma ^{(0)}}\frac{1}{2z}\left(\frac{2}{3}f-\frac{5}{24}R\right).
\end{aligned}
\end{equation}
Together with the first equality in \eqref{eq:Bianchi}, this  further implies
\begin{equation}
A'_{\text{reg, }O(z^{-1})}=\frac{1}{16z}\partial _kR.
\end{equation}
Therefore we have
\begin{align}
g_{(0)ij}(\nabla _z X^i_{(4)})k^j =-\frac{z^3}{16}\partial _kR,\\
g_{(0)ij}X^i_{(4)}k^j =-\frac{z^4}{64}\partial _kR,\\
g_{(0)ij}(\nabla _\lambda X^i_{(4)})k^j =-\frac{z^4}{64}\partial _k^2R. \label{g0X4kd=5}
\end{align}
Plugging equation \eqref{g0X4kd=5} and \eqref{eq:g4kk5d} into \eqref{eq:inequality5d}, we find as expected that the $O(z^{4})$ terms cancel:
\begin{equation}\label{Oz4}
g_{(4)ij} k^ik^j + 2 g_{(0)ij} (\nabla_{\lambda}X^i_{(4)})k^j = 0.
\end{equation}
Combining the above results yields
\begin{align}
A'_\text{ren}= & A'_\text{reg, finite}+A'_\text{ct, finite}
=  -\frac{1}{z^4}g_{(0)ij}(\nabla _zX^i_{(5)})k^j+0,
\end{align}
which implies
\begin{align}\label{eq:renArea}
	g_{(0)ij}X^i_{(5)}k^j = -\frac{z^{5}}{5} A'_{\rm ren}.
\end{align}
Using (\ref{eq:inequality5d}), (\ref{eq:g5kk}), (\ref{Oz4}), (\ref{eq:renArea}), we thus obtain the renormalized QNEC
\begin{align}
	0 \leq T_{kk} - \frac{1}{8\pi G} A''_{\rm ren}.
\end{align}

\subsection{Killing horizons}\label{sec:killingproof}

As noted in section \ref{sec:schemedep}, a result of \cite{Wall:2015raa} implies the QNEC to be scheme-independent on any bifurcate Killing horizon.  So for any $d$ we would expect the holographic argument to yield a finite renormalized QNEC in this case as well.

In particular, we consider a boundary metric \(g_{(0)ij}\) with a bifurcate Killing horizon \(H_{(0)}\) generated by the Killing vector \(\xi_{(0)}^{i}\), i.e.
\begin{align}
	\lie_{\xi}g_{(0)ij} = 0
\end{align}
We evaluate the QNEC for deformations generated by $k^i \propto \xi^{i}_{(0)}$ acting on a cut \(\partial A\) of \(H_{(0)}\).

The critical fact is that, as explained above, the possible (finite or divergent) corrections to \eqref{flatchi^{2}} are of the form $Z_{ij} k^i k^j$ for some smooth (covariant) geometric tensor $Z_{ij}$ built from the spacetime metric $g_{(0)ij}$, the extrinsic curvature $K^i_{~jk}$, the projector $h_{(0)i}{}^j$ onto $\partial A$, and their derivatives.    Since $\xi_{(0)}^i$ generates a symmetry, the quantity $Z_{ij} \xi_{(0)}^i \xi_{(0)}^j$ is some constant $C$ along the flow generated by $\xi_{(0)}^i$, and thus along each generator.  But $\xi_{(0)}^i = fk^i$ for some scalar function $f$ that vanishes on the bifurcation surface.  The fact that $Z_{ij} k^i k^j = f^{-2} C$ must be smooth and thus finite at the bifurcation surface then forces $C=0$, so all possible corrections to \eqref{flatchi^{2}} in fact vanish. Note that this argument relies only on the general form of the Fefferman-Graham expansion and on the HRT action \eqref{eq:HRTaction}. It does not depend on the detailed equations of motion.

\section{Discussion}
\label{sec:discussion}
We have investigated the QNEC in curved space by analyzing the scheme-independence of the QNEC and its validity in holographic field theories.  For $d\le 3$, for arbitrary backgroud metric we found that the QNEC \eqref{QNECintro} is naturally finite and  independent of renormalization scheme for points $p$ and null congruences $N$ for which the expansion $\theta$ vanishes at $p$.  It is interesting that this condition is weaker than the local stationarity assumption ($\theta |_p= \dot{\theta} |_p=0, \sigma_{ab}|_p=0$) under which the QNEC was previously proposed to hold, and it is in particular weaker than the conditions under which it can be derived from the quantum focusing conjecture \cite{Bousso:2015mna}. But for $d=4,5$ we require local stationarity as well as the vanishing of additional derivatives as in \eqref{eq:highD}, as well as a dominant energy condition \eqref{gDEC}. Under the above conditions, we also showed the universal sector of leading-order holographic theories to satisfy a finite renormalized QNEC.

The success of this derivation for $d\le 3$ (using only $\theta=0$) suggests that the QNEC may hold for general field theories in contexts where it cannot be derived from the quantum focusing conjecture (QFC).  If so, it would be incorrect to think of the QFC as being more fundamental than the QNEC;  the QNEC seems to have a life of its own. 

For $d\ge6$ we argued these properties to generally fail even for weakly isolated horizons (where all derivatives of $\theta, \sigma_{ab}$ vanish) satisfying the dominant energy condition, though they do hold on Killing horizons. The issue in \(d=6\) is that finite counter-terms in the effective action can contain derivatives of the Riemann tensor, and that these terms change the definition of the entropy and stress tensor in such a way that the combination entering the QNEC is not invariant.

Our $d=5$ argument for scheme-independence required the conditions \eqref{eq:highD} and \eqref{gDEC}, while for $d=4$ we need only \eqref{gDEC} and the first line of \eqref{eq:highD}. It certainly appears that local stationarity is not itself sufficient for the four-derivative counter-terms in \eqref{eq:d45}, this has been shown (using \cite{Fu:2017lps} and in footnote \ref{Fufoot}) only for $d\ge 5$ in the case of the Gauss-Bonnet term. For $d=4$ the Gauss-Bonnet term gives $\Delta Q=0$. Terms involving only the scalar curvature (i.e., the $R^2$ term) are easily handled by changing conformal frame to write the theory as Einstein-Hilbert gravity coupled to a scalar field \cite{Wands:1993uu}. So if the QNEC is invariant under the remaining $R_{ab}R^{ab}$ term, one would expect a useful QNEC (and thus perhaps also a useful quantum focussing condition (QFC)) to hold in $d=4$ as well.

As explained in \cite{Bousso:2015mna}, the QNEC implies the perturbative semi-classical generalized second law (GSL) of thermodynamics at first non-trivial order in the gravitational coupling $G$.   A consequence of our work is a thus proof of the (first-order) semi-classical GSL on causal horizons satisfying the conditions above, and in particular on general causal horizons for $d\le 3$.  Even at this order, this is the first GSL proof valid when the null congruence $N$ does not reduce to a Killing horizon in the background.

Now, as described in footnote \ref{Fufoot}, it was recently shown in \cite{Fu:2017lps} that for $d \ge 5$ the QNEC generally fails to be scheme-independent when the change of renormalization scheme entails the addition of a Gauss-Bonnet term to the action, and furthermore that the associated change $\Delta Q$ can have either sign. It then follows that (for theories that require a Gauss-Bonnet counter-term) a renormalized QNEC cannot hold in general renormalization schemes as, if one finds a finite $Q \ge 0$ with some scheme, we may always change the scheme to add a Gauss-Bonnet term so that $Q_{\rm modified} = Q + \Delta Q_{\rm GB} <  0$.

In a rather different direction, it was recently noted \cite{Akers:2016ugt,Koeller:2017njr} that on Killing horizons the quantum null energy condition is related to a property of the relative entropy \(S(\rho||\sigma)\) between an arbitrary state \(\rho\) and the vacuum state \(\sigma\):
\begin{align}
	0 \leq T_{kk} - \frac{\hbar}{2\pi} S'' = S(\rho||\sigma)''
\end{align}
In this equation, the derivatives of the relative entropy are the same type of local derivatives with respect to null deformations of the region that appear in \(S''\). (This ``concavity'' property is about the {\it second} derivative, while the well-known monotonicity of relative entropy bounds the {\it first} derivative, \(S(\rho||\sigma)' \leq 0\).)  While we have seen that the QNEC is not always well defined or true in curved space, the relative entropy is known to be scheme-independent. It would thus be interesting to understand if  an inequality of the form \(S(\rho||\sigma)'' \geq 0\) for appropriate $\sigma$ might hold more generally, perhaps even in cases where the QNEC fails.  One might also investigate whether, without introducing any smearing, this could lead to a new conjecture for theories with dynamical gravity that could replace the quantum focussing condition  \cite{Bousso:2015mna} and which might hold even when the original QFC is violated \cite{Fu:2017lps}.

Finally, we comment on the relation of the QNEC to the observation of \cite{Leichenauer:2017bmc} that at least the QFC violation of \cite{Fu:2017lps} can be avoided by taking the QFC to apply only to suitably smooth variations of the generalized entropy.  One might then ask if such a ``smeared QFC'' could lead to a suitably smeared version of the QNEC that would hold even when the original QNEC fails.    However, the original QFC reduces to the QNEC only at locally stationary points where $\theta |_p= \dot{\theta}|_p= \sigma_{ab}|_p=0$. And the point of the averaging in \cite{Leichenauer:2017bmc} is precisely that, when $d > 3$ and $k^a h_e^{~b}h_f^{~c}h_g^{~d}R_{abcd} \neq 0$ (where $k^a$ and $h_{a}{}^{b}$ are respectively the null normal vector and the projector onto the chosen cut of the null congruence $N$), the locally stationary condition can hold only a set of measure zero.  Dimensional analysis then shows that smearing out the QFC on scales long compared to the cutoff leads manifestly non-positive QFC contributions from violations of local stationarity to swamp those from failures of the QNEC.  In other words, any QNEC-like inequality is irrelevant to the smeared QFC of \cite{Leichenauer:2017bmc} unless
\begin{equation}
\label{QNECcond}
d\le 3\text{ or }k^a h_e^{~b}h_f^{~c}h_g^{~d}R_{abcd}=0,
\end{equation}
so that only under one of these conditions could any QNEC be derived from this smeared QFC.  We therefore suspect that these are the most general conditions under which any QNEC could possibly hold, and the analysis of section \ref{scheme-indep}  suggests that further conditions are likely required at least for $d\ge 6$.  Indeed, as shown in appendix \ref{appendix:identities}, the condition \eqref{QNECcond} follows from \eqref{eq:highD}, and \eqref{gDEC},  and then from \cite{Fu:2017lps} one sees that it suffices to avoid the $d\ge 5$ QNEC violation associated with the Gauss-Bonnet term.  However, it remains to further investigate the effect of the $R_{ab}R^{ab}$ counter-term for both $d=4,5$ because \eqref{QNECcond} does not appear to guarantee the existence of a null congruence $N$ satisfying the sufficient conditions which we used in this paper.  Similar comments must also apply to the proposed ``quantum dominant energy condition'' of \cite{Wall:2017blw}, which reduces to the QNEC when considering pairs of variations that act in the same null direction.

\section*{Acknowledgements}
It is a pleasure to thank Chris~Akers, Raphael~Bousso, Venkatesh~Chandrasekaran, Xi~Dong, Stefan~Leichenauer, Adam~Levine, Rong-Xin~Miao and Arvin~Moghaddam for discussions. JK would especially like to thank Stefan~Leichenauer for many useful discussions on related ideas. ZF and DM were supported in part by the Simons Foundation and by funds from the University of California. JK was supported in part by the Berkeley Center for Theoretical Physics, by the National Science Foundation (award numbers 1521446, and 1316783), by FQXi, and by the US Department of Energy under contract DE-AC02-05CH11231.

\appendix

\section{Non-Expanding Horizons and Weakly Isolated Horizons}
\label{appendix:identities}

In this appendix, we study properties of non-expanding horizons and weakly isolated horizons. First, we provide geometric identities for non-expanding null surfaces. We use $k^a$ to indicate the generator of a null surface which satisfies the null condition $k_ak^a=0$ and the geodesic equation $k^b\nabla _bk^a=0$. We introduce an auxiliary null vector field $l^a$ satisfying $l_al^a=0$, $k_al^a=-1$, and $\pounds _kl^a=0$. The transverse metric of the null surface is given by $h_{ab} = g_{ab}+k_al_b+l_ak_b$. To make the notation more precise, we use the sign ``$\heq $'' to denote ``equal on the horizon'' in this appendix and in appendix \ref{appendix:schemeindependence}.

A non-expanding null surface is defined by
\begin{align}\label{Bhat}
h_{a}^{~c}h_{b}^{~d} \nabla_{c} k_{d} \heq 0.
\end{align}
Substituting the definition of $h_{ab}$ into the above equation, we obtain
\begin{align}\label{nablak}
\nabla_{a}k_{b} \heq L_ak_b+k_aR_b+Bk_ak_b,
\end{align}
where $L_a\equiv -l^d\nabla_{a} k_{d}$, $R_b\equiv -l^c\nabla_{c} k_{b}$, and $B\equiv -l^cl^d\nabla_{c} k_{d}$. $L^a$, $R^a$ and $B$ satisfy relations $L_ak^a=0$, $R_bk^b=0$, and $L_al^a=B=R_bl^b$. Furthermore, there is
\begin{equation}
\nabla_{a}k^{a} \heq 0.
\end{equation}
The extrinsic curvature, defined by
\begin{equation}
\label{eq:DefExtrCurv}
K^{c}_{~ab} := -h_{a}^{~d}h_{b}^{~e}\nabla _dh_{e}^{~c},
\end{equation}
of a non-expanding null surface can always be written as
\begin{equation}
K^c_{~ab}\heq k^cA_{ab},
\end{equation}
where $A_{ab}\equiv -h^{~d}_a\nabla _dl_b+L_al_b+Bk_al_b$.

On a non-expanding null surface, the Riemann tensor contracting with a $k^a$ can be written in terms of $L^a$, $R^a$, and $B$ as
\begin{equation}
\label{eq:Rabck}
\begin{aligned}
R_{abck}&\equiv 2\nabla _{[a}\nabla _{b]}k_c
\\
&\heq 2k_c\nabla _{[a}L_{b]}+2R_ck_{[a}R_{b]}+2k_{[b}\nabla _{a]}R_c+2k_ck_{[b}\nabla _{a]}B+2Bk_ck_{[a}R_{b]}+2Bk_{[b}L_{a]}k_c.
\end{aligned}
\end{equation}
From the above equation, one immediately obtains
\begin{align}
\label{eq:NEHRakck}
R_{akck}&\heq -L_{b}k_ck_{a}R^b-k^bk_c\nabla _{b}L_{a}-k^bk_{a}\nabla _{b}R_c-k^bk_ck_{a}\nabla _{b}B, \\
R_{bk}&\heq k^a\nabla _{a}L_{b}+k_{b}\nabla _{a}R^a+k_{b}L^aR_{a}+k_{b}\partial _{k}B, \\
\label{eq:NEHRkk}
R_{kk}&\heq 0.
\end{align}

The above results are for non-expanding horizons in general. By further imposing the dominant energy condition \eqref{gDEC}, one observes special properties of weakly isolated horizons. First, by noticing that equation \eqref{eq:NEHRkk} implies that vector $R^a_{~b}k^b$ must be a vector tangent to the horizon and equation \eqref{gDEC} implies that $R^a_{~b}k^b$ can be written as a term proportional to $k^a$ plus a causal vector, one concludes that for weakly isolated horizons there is
\begin{equation}
\label{eq:Rka}
R_{ab}k^b\heq fk_a.
\end{equation}
Second, we argue that on weakly isolated horizons the Weyl tensor is Petrov type II. To prove this, we only need to show that for weakly isolated horizons there is
\begin{align}
\label{eq:PetrovI}
C_{abcd}k^ak^ch^b_{~e}h^d_{~f}&\heq 0,\\
\label{eq:PetrovII}
C_{abcd}k^ah^b_{~e}h^c_{~f}h^d_{~g}&\heq 0.
\end{align}
Equation \eqref{eq:Rka} and equation
\begin{equation}
\label{eq:WeylDecomposition}
R_{abcd}=C_{abcd}+\frac{2}{d-2}\left(g_{a[c}R_{d]b}-g_{b[c}R_{d]a}\right)-\frac{2}{\left(d-1\right)\left(d-2\right)}Rg_{a[c}g_{d]b}
\end{equation}
together imply that
\begin{align}
\label{Rkbkd}
C_{abcd}k^ak^ch^b_{~e}h^d_{~f}&\heq R_{abcd}k^ak^ch^b_{~e}h^d_{~f},\\
\label{Rkabc}
C_{abcd}k^ah^b_{~e}h^c_{~f}h^d_{~g}&\heq R_{abcd}k^ah^b_{~e}h^c_{~f}h^d_{~g}.
\end{align}
Equation \eqref{eq:Rabck} implies that the right-hand side of equation \eqref{Rkabc} vanishes and equation \eqref{eq:NEHRakck} implies that the right-hand side of equation \eqref{Rkbkd} vanishes. Therefore, equations \eqref{eq:PetrovI} and \eqref{eq:PetrovII} do hold and the Wyel tensor is indeed Petrov type II on weakly isolated horizons. This leads us to conclude that on weakly isolated horizons we have
\begin{equation}
C_{abcd}k^bk^d\heq \hat{\zeta }k_ak_c,
\end{equation}
which, together with \eqref{eq:WeylDecomposition} and \eqref{eq:Rka}, further leads to
\begin{equation}
\label{eq:Rakck}
R_{abcd}k^bk^d\heq \zeta k_ak_c.
\end{equation}
The contracted Bianchi identity $\nabla _aR_{bcd}^{~~~a}+\nabla _bR_{cd}-\nabla _cR_{bd}=0$ provides a relationship among $\zeta $, $f$, and the spacetime scalar curvature $R$:
\begin{equation}
\label{eq:WIHBianchi}
\frac{1}{2}\partial_kR\heq \partial_kf\heq -\partial_k\zeta .
\end{equation}

These features guide one to conclude $k^a\nabla _{a}L_{b}\propto k_b$ and $k^b\nabla _b R_c\propto k_c$. Therefore, on weakly isolated horizons, the Riemann tensor $R_{abck}$ can be written as $R_{abck}\heq k_c\tilde{A}_{ab}+k_{[a}\tilde{B}_{b]c}$, where $\tilde{A}_{ab}$ is antisymmetric and satisfies $k^a\tilde{A}_{ab}\propto k_b$ and $\tilde{B}_{ab}$ satisfies $k^a\tilde{B}_{ab}\propto k_b$ and $k^b\tilde{B}_{ab}\propto k_a$. This guarantees more proportionl relations, including $R^{ac}R_{abck}\propto k_b$, $R_k^{~bcd}R_{cdbe}\propto k_e$, $R_{kec}^{~~~f}R_{afk}^{~~~e}\propto k_ak_c$, and $R^{ab}_{~~ck}R_{abdk}\propto k_ck_d$.

These relations are crucial to the scheme-independence of QNEC for $d=4,5$.   However, in those cases we do not require them to hold on the entire horizon, but only to the appropriate order in $\lambda$ about the point $p$. As a result, it suffices to impose only \eqref{eq:highD}.


\section{Scheme Independence of the QNEC}
\label{appendix:schemeindependence}

In this appendix, we derive the scheme independence of the QNEC for four-derivative counter-terms and six-derivative counter-terms built from arbitrary polynomial contractions of the Riemann tensor. Equations \eqref{nablak}, \eqref{eq:Rka}, and \eqref{eq:Rakck} are our three inputs. As we have mentioned at the end of appendix \ref{appendix:identities}, with these three assumptions the Riemann tensor $R_{abck}$ can be written in the form $R_{abck}\heq k_c\tilde{A}_{ab}+k_{[a}\tilde{B}_{b]c}$, where $\tilde{A}_{ab}$ is antisymmetric and satisfies $k^a\tilde{A}_{ab}\propto k_b$ and $\tilde{B}_{ab}$ satisfies $k^a\tilde{B}_{ab}\propto k_b$ and $k^b\tilde{B}_{ab}\propto k_a$. Moreover, the contracted Bianchi identity implies that equation \eqref{eq:WIHBianchi} holds on the horizon. We introduce notations $g^\perp _{ab}\equiv -k_al_b-l_ak_b$ and $\epsilon _{ab}\equiv -k_al_b+l_ak_b$, where the meaning of $l^a$ has been explained at the beginning of appendix \ref{appendix:identities}.

In principle, to calculate the gravitational entropy associated with the type of counter-terms considered in this appendix, we need to use Dong's entropy formula \cite{Dong:2013qoa}. However, it is easy to see that the $S'_{\text{ct}}$ from Dong's entropy formula reduces to that computed from Wald's formula when \eqref{eq:highD} and \eqref{gDEC} hold. Therefore, in this appendix, we compute the gravitational entropy from the formula
\begin{equation}
S'_{\text{ct}}=\partial _k\left(-2\pi \epsilon _{ab}\epsilon _{cd}\frac{\partial L_{\text{ct}}}{\partial R_{abcd}}\right).
\end{equation}
Furthermore, the $kk$-component of the stress tensor associated with these counter-terms is given by the standard functional derivative formula
\begin{equation}
T_{kk\text{, ct}}=k^bk^d\frac{-2}{\sqrt{-g}}\frac{\delta I_{\text{ct}}}{\delta g^{bd}}.
\end{equation}
The change of the QNEC associated with these counter-terms is thus given by
\begin{equation}
Q_{\text{ct}}=T_{kk\text{, ct}}-\frac{1}{2\pi }S''_{\text{ct}}.
\end{equation}

\subsection{Four-derivative counter-terms}
We now consider the three possible four-derivative counter-terms,
\begin{equation}
I_1=\int d^dx\sqrt{-g}R^{abcd}R_{abcd},~I_2=\int d^dx\sqrt{-g}R^{ab}R_{ab}, \ \text{and} \ ~I_3=\int d^dx\sqrt{-g}R^2.
\end{equation}

For the counter-term $I_3$, the entropy is
\begin{equation}
S'_{\text{ct3}}=\partial _k\left(8\pi R\right),
\end{equation}
so that
\begin{equation}
\frac{1}{2\pi }S_{\text{ct3}}''=4 \partial _k^2R.
\end{equation}
To compute the stess tensor term we write
\begin{equation}
\delta I_3=\int d^dx\sqrt{-g}2R\left(\nabla ^d\nabla ^b\delta g_{bd}-g^{bd}\nabla ^c\nabla _c\delta g_{bd}\right),
\end{equation}
\begin{equation}
T_{kk\text{, ct3}}=k^bk^d\frac{-2}{\sqrt{-g}}\frac{\delta I_3}{\delta g^{bd}}=4k^bk^d\nabla _b\nabla _dR=4\partial _k^2R,
\end{equation}
so the QNEC remains unchanged
\begin{equation}
\Delta Q_{\text{ct3}}=T_{kk\text{, ct3}}-\frac{1}{2\pi }S_{\text{ct3}}''=0.
\end{equation}

For $I_2$, we find
\begin{equation}
S'_{\text{ct2}}=\partial _k\left(4\pi g^{\perp ab}R_{ab}\right)=\partial _k\left(-8\pi R_{kl}\right)=\partial _k\left(8\pi f\right),
\end{equation}
\begin{equation}
\frac{1}{2\pi }S_{\text{ct2}}''=4 \partial _k^2f,
\end{equation}
and
\begin{equation}
\delta I_2=\int d^dx\sqrt{-g}2R^{ab}\left(-\frac{1}{2}g^{cd}\nabla _a\nabla _b\delta g_{cd}-\frac{1}{2}g^{cd}\nabla _c\nabla _d\delta g_{ab}+g^{cd}\nabla _c\nabla _b\delta g_{ad}\right),
\end{equation}
\begin{equation}
T_{kk\text{, ct2}}=k^ak^b\frac{-2}{\sqrt{-g}}\frac{\delta I_2}{\delta g^{ab}}=-2k^ak^b\nabla _c\nabla ^cR_{ab}+4k^ak^b\nabla _c\nabla _bR_a^{~c}=4\partial _k^2f.
\end{equation}
So again we find
\begin{equation}
\Delta Q_{\text{ct2}}=T_{kk\text{, ct2}}-\frac{1}{2\pi }S_{\text{ct2}}''=0.
\end{equation}

For $I_1$, we have
\begin{equation}
S'_{\text{ct1}}=\partial _k\left(8\pi g^{\perp ac}g^{\perp bd}R_{abcd}\right)=\partial _k\left(-16\pi R_{lklk}\right)=\partial _k\left(-16\pi \zeta \right),
\end{equation}
\begin{equation}
\frac{1}{2\pi }S_{\text{ct1}}''=-8 \partial _k^2\zeta ,
\end{equation}
and
\begin{equation}
\delta I_1=\int d^dx\sqrt{-g}2R^{abcd}\left(-2\nabla _a\nabla _c\delta g_{bd}\right),
\end{equation}
\begin{equation}
T_{kk\text{, ct1}}=k^bk^d\frac{-2}{\sqrt{-g}}\frac{\delta I_2}{\delta g^{bd}}=-8k^bk^d\nabla ^c\nabla ^aR_{abcd}=-8\partial _k^2\zeta ,
\end{equation}
and again
\begin{equation}
\Delta Q_{\text{ct1}}=T_{kk\text{, ct1}}-\frac{1}{2\pi }S_{\text{ct1}}''=0.
\end{equation}

These results are summarized in table \ref{tb:4dcounterterm}.

\subsection{Six-derivative counter-terms}
We now consider six-derivative counter-terms built from arbitrary polynomial contractions of the Riemann tensor. These terms are listed in \cite{Oliva:2010zd}. They are
\begin{equation}
\begin{aligned}
I_1=\int d^dx\sqrt{-g} R^{abcd}R_{cdef}R^{ef}_{~~ab}\text{, }&I_2=\int d^dx\sqrt{-g} R^{ab}_{~~cd}R^{ce}_{~~bf}R^{df}_{~~ae}, \\
I_3=\int d^dx\sqrt{-g} R^{abcd}R_{cdbe}R^{e}_{~a}\text{, }&I_4=\int d^dx\sqrt{-g} RR^{abcd}R_{abcd}, \\
I_5=\int d^dx\sqrt{-g} R^{abcd}R_{ac}R_{bd}\text{, }&I_6=\int d^dx\sqrt{-g} R^{ab}R_{bc}R^{c}_{~a}, \\
I_7=\int d^dx\sqrt{-g} RR^{ab}R_{ab}\text{, }&I_8=\int d^dx\sqrt{-g} R^3.
\end{aligned}
\end{equation}

For $I_8$, we find
\begin{equation}
S'_{\text{ct8}}=\partial _k\left(12\pi R^2\right),
\end{equation}
\begin{equation}
\frac{1}{2\pi }S_{\text{ct8}}''=6 \partial _k^2R^2,
\end{equation}
and
\begin{equation}
\delta I_8=\int d^dx\sqrt{-g}3R^2\left(\nabla ^d\nabla ^b\delta g_{bd}-g^{bd}\nabla ^c\nabla _c\delta g_{bd}\right),
\end{equation}
\begin{equation}
T_{kk\text{, ct8}}=k^bk^d\frac{-2}{\sqrt{-g}}\frac{\delta I_3}{\delta g^{bd}}=k^bk^d(6)\nabla _b\nabla _dR^2=6\partial _k^2R^2,
\end{equation}
and thus
\begin{equation}
\Delta Q_{\text{ct8}}=T_{kk\text{, ct8}}-\frac{1}{2\pi }S_{\text{ct8}}''=0.
\end{equation}

For $I_7$, we find
\begin{equation}
S'_{\text{ct7}}=\partial _k\left[-2\pi \left(-2R^{ab}R_{ab}-4fR\right)\right],
\end{equation}
\begin{equation}
\frac{1}{2\pi }S_{\text{ct7}}''=2\partial _k^2\left(R^{ab}R_{ab}\right)+4\partial _k^2\left(fR\right),
\end{equation}
and
\begin{equation}
\begin{aligned}
\delta I_7=&\int d^dx\sqrt{-g}R^{ab}R_{ab}\left(\nabla ^d\nabla ^c\delta g_{cd}-g^{cd}\nabla ^e\nabla _e\delta g_{cd}\right)\\
&+\int d^dx\sqrt{-g}2RR^{ab}\left(-\frac{1}{2}g^{cd}\nabla _a\nabla _b\delta g_{cd}-\frac{1}{2}g^{cd}\nabla _c\nabla _d\delta g_{ab}+g^{cd}\nabla _c\nabla _b\delta g_{ad}\right),
\end{aligned}
\end{equation}
\begin{equation}
\begin{aligned}
T_{kk\text{, ct7}}&=k^ck^d\frac{-2}{\sqrt{-g}}\frac{\delta I_7}{\delta g^{cd}}\\
&=2k^ck^d\nabla _c\nabla _d\left(R^{ab}R_{ab}\right)-2k^ck^d\nabla _e\nabla ^e\left(RR_{cd}\right)+4k^ck^d\nabla _b\nabla _d\left(RR_c^{~b}\right)\\
&=2\partial _k^2\left(R^{ab}R_{ab}\right)+4\partial _k^2\left(Rf\right),
\end{aligned}
\end{equation}
and thus
\begin{equation}
\Delta Q_{\text{ct7}}=T_{kk\text{, ct7}}-\frac{1}{2\pi }S_{\text{ct7}}''=0.
\end{equation}

For $I_6$, we have
\begin{equation}
S'_{\text{ct6}}=\partial _k\left(6\pi R_{bc}R^c_{~a}g^{\perp ab}\right)=-\partial _k\left(12\pi R_{kc}R^c_{~l}\right)=\partial _k\left(12\pi f^2\right),
\end{equation}
\begin{equation}
\frac{1}{2\pi }S_{\text{ct6}}''=6\partial _k^2\left(f^2\right),
\end{equation}
and
\begin{equation}
\delta I_6=\int d^dx\sqrt{-g} 3R^{be}R_e^{~a}\left(-\frac{1}{2}g^{cd}\nabla _a\nabla _b\delta g_{cd}-\frac{1}{2}\nabla ^c\nabla _c\delta g_{ab}+\nabla ^d\nabla _b\delta g_{ad}\right),
\end{equation}
\begin{equation}
\begin{aligned}
T_{kk\text{, ct6}}&=k^ak^b\frac{-2}{\sqrt{-g}}\frac{\delta I_6}{\delta g^{ab}}\\
&=-3k^ak^b\nabla _c\nabla ^c\left(R_{be}R^e_{~a}\right)+6k^ak^b\nabla _c\nabla _b\left(R^{ce}R_{ea}\right)\\
&=6\partial _k^2\left(f^2\right),
\end{aligned}
\end{equation}
and so
\begin{equation}
\Delta Q_{\text{ct6}}=T_{kk\text{, ct6}}-\frac{1}{2\pi }S_{\text{ct6}}''=0.
\end{equation}

To deal with counter-term $I_5$, not first that $R_{kb}\heq fk_b$ implies $R^{ac}R_{abck}\heq \tilde{f}k_b$ and that contracting both sides with $l^b$ then gives the function $\tilde{f}\heq -R^{ac}R_{alck}$. Thus we find
\begin{equation}
R^{ac}R_{abck}\heq -k_bR^{ac}R_{alck}.
\end{equation}
We may now compute
\begin{equation}
\begin{aligned}
S'_{\text{ct5}}&=\partial _k\left(2\pi g^\perp _{~[a\vert c\vert }g^\perp _{~b]d}R^{ac}R^{bd}+4\pi g^\perp _{~bd}R^{abcd}R_{ac}\right)\\
&=\partial _k\left(4\pi f^2-8\pi R^{akcl}R_{ac}\right),
\end{aligned}
\end{equation}
\begin{equation}
\frac{1}{2\pi }S_{\text{ct5}}''=2\partial _k^2\left(f^2\right)-4\partial _k^2\left(R^{akcl}R_{ac}\right),
\end{equation}
and
\begin{equation}
\begin{aligned}
\delta I_5=&\int d^dx\sqrt{-g}R^{a[c}R^{\vert b\vert d]}\left(-2\nabla _a\nabla _c\delta g_{bd}\right) \\
&+\int d^dx\sqrt{-g}R^{abcd}2R_{ac}\left(-\frac{1}{2}g^{ef}\nabla _b\nabla _d\delta g_{ef}-\frac{1}{2}\nabla _e\nabla ^e\delta g_{bd}+\nabla ^f\nabla _d\delta g_{bf}\right),
\end{aligned}
\end{equation}
\begin{equation}
\begin{aligned}
T_{kk\text{, ct5}}= & k^bk^d\frac{-2}{\sqrt{-g}}\frac{\delta I_5}{\delta g^{bd}} \\
= & -4k^bk^d\nabla ^c\nabla ^aR_{a[c}R_{\vert b\vert d]}-2k^bk^d\nabla ^e\nabla _e\left(R_{abcd}R^{ac}\right)+4k^bk^f\nabla ^d\nabla _f\left(R_{abcd}R^{ac}\right) \\
= & 2k^bk^d\nabla ^c\nabla ^aR_{ad}R_{bc}+4k^bk^f\nabla ^d\nabla _f\left(R_{abcd}R^{ac}\right) \\
= & 2\partial _k^2\left(f^2\right)-4\partial _k^2\left(R_{akcl}R^{ac}\right),
\end{aligned}
\end{equation}
which together imply
\begin{equation}
\Delta Q_{\text{ct5}}=T_{kk\text{, ct5}}-\frac{1}{2\pi }S_{\text{ct5}}''=0.
\end{equation}

For $I_4$, we find
\begin{equation}
S'_{\text{ct4}}=\partial _k\left[-2\pi \left(-2R^{abcd}R_{abcd}+8\zeta R\right)\right],
\end{equation}
\begin{equation}
\frac{1}{2\pi }S_{\text{ct4}}''=2\partial _k^2\left(R^{abcd}R_{abcd}\right)-8\partial _k^2\left(\zeta R\right),
\end{equation}
and
\begin{equation}
\begin{aligned}
\delta I_4=&\int d^dx\sqrt{-g}R^{abcd}R_{abcd}\left(\nabla ^f\nabla ^e\delta g_{ef}-g^{ef}\nabla ^g\nabla _g\delta g_{ef}\right)\\
&+\int d^dx\sqrt{-g}2RR^{abcd}\left(-2\nabla _a\nabla _c\delta g_{bd}\right),
\end{aligned}
\end{equation}
\begin{equation}
\begin{aligned}
T_{kk\text{, ct4}}&=k^bk^d\frac{-2}{\sqrt{-g}}\frac{\delta I_4}{\delta g^{bd}}\\
&=2k^ek^f\nabla _e\nabla _f\left(R^{abcd}R_{abcd}\right)-8k^bk^d\nabla ^c\nabla ^a\left(RR_{abcd}\right)\\
&=2\partial _k^2\left(R^{abcd}R_{abcd}\right)-8\partial _k^2\left(R\zeta \right),
\end{aligned}
\end{equation}
so that
\begin{equation}
\Delta Q_{\text{ct4}}=T_{kk\text{, ct4}}-\frac{1}{2\pi }S_{\text{ct4}}''=0.
\end{equation}

For counter-term $I_3$, note that $R_{akck}=\zeta k_ak_c$ implies $R_k^{~bcd}R_{cdbe}=\tilde{\zeta }k_e$. Contracting both sides with $l^e$ then gives the function $\tilde{\zeta }=-R_k^{~bcd}R_{cdbl}$. Thus we find
\begin{equation}
R_k^{~bcd}R_{cdbe}=-k_eR_k^{~bcd}R_{cdbl}.
\end{equation}
We may now compute
\begin{equation}
\begin{aligned}
S'_{\text{ct3}}= & \partial _k\left(-8\pi R^{cdbe}R_e^{~a}g^\perp _{~[a|c|}g^\perp _{~b]d}+2\pi R^{abcd}R_{cdb}^{~~~e}g^\perp _{~ea}\right)\\
= & \partial _k\left(-16\pi R_{lkle}R_k^{~e}-4\pi R_k^{~bcd}R_{cdbl}\right)\\
= & \partial _k\left(-16\pi \zeta f-4\pi R_k^{~bcd}R_{cdbl}\right),
\end{aligned}
\end{equation}
\begin{equation}
\frac{1}{2\pi }S_{\text{ct3}}''=-8\partial _k^2\left(\zeta f\right)-2\partial _k^2\left(R_k^{~bcd}R_{cdbl}\right),
\end{equation}
and
\begin{equation}
\begin{aligned}
\delta I_3= & \int d^dx\sqrt{-g}2R^{cdbe}R_e^{~a}\left(-2\nabla _a\nabla _c\delta g_{bd}\right)\\
& + \int d^dx\sqrt{-g}R^{abcd}R_{cdb}^{~~~e}\left(-\frac{1}{2}g^{fg}\nabla _a\nabla _e\delta g_{fg}-\frac{1}{2}\nabla _f\nabla ^f\delta g_{ea}+\nabla ^f\nabla _e\delta g_{af}\right),
\end{aligned}
\end{equation}
\begin{equation}
\begin{aligned}
T_{kk\text{, ct3}}= & k^bk^d\frac{-2}{\sqrt{-g}}\frac{\delta I_3}{\delta g^{bd}} \\
= & -8k_bk_d\nabla _c\nabla _a\left(R^{cdbe}R_e^{~a}\right)-k^ek_a\nabla ^f\nabla _f\left(R^{abcd}R_{cdbe}\right)\\
& 2k_ak^f\nabla ^e\nabla _f\left(R^{abcd}R_{cdbe}\right)\\
= & -8\partial _k^2\left(\zeta f\right)-2\partial _k^2\left(R_k^{~bcd}R_{cdbl}\right).
\end{aligned}
\end{equation}
Putting these together yields
\begin{equation}
\Delta Q_{\text{ct3}}=T_{kk\text{, ct3}}-\frac{1}{2\pi }S_{\text{ct3}}''=0.
\end{equation}

For counter-term $I_2$, notice that $R_{akck}=\zeta k_ak_c$ implies $R^e_{~afk}R^{~~f}_{ec~k}=\bar \zeta k_ak_c$. Contracting both sides with $l^al^c$ gives the function $\bar \zeta =R^e_{~lfk}R^{~~f}_{el~k}$. One then finds
\begin{equation}
R^e_{~afk}R^{~~f}_{ec~k}=k_ak_cR^e_{~lfk}R^{~~f}_{el~k}.
\end{equation}
With this in hand, we calculate
\begin{equation}
\begin{aligned}
S'_{\text{ct2}}= & \partial _k\left(-12\pi g^{\perp [a}_{~~~c}g^{\perp b]}_{~~~d}R^{ce}_{~~bf}R^{df}_{~~ae}\right)\\
= & \partial _k\left[-12\pi \left(R_{kel}^{~~~f}R_{lfk}^{~~~e}-R_{lel}^{~~~f}R_{kfk}^{~~~e}\right)\right]\\
= & \partial _k\left[-12\pi \left(R_{kel}^{~~~f}R_{lfk}^{~~~e}-\zeta ^2\right)\right],
\end{aligned}
\end{equation}
\begin{equation}
\frac{1}{2\pi }S_{\text{ct2}}''=-6\partial _k^2\left(R_{kel}^{~~~f}R_{lfk}^{~~~e}-\zeta ^2\right),
\end{equation}
and
\begin{equation}
\delta I_2=\int d^dx\sqrt{-g}3R^{[c}_{~~egf}R^{d]fae}g^{gb}\left(-2\nabla _a\nabla _c\delta g_{bd}\right),
\end{equation}
\begin{equation}
\begin{aligned}
T_{kk\text{, ct2}}= & k^bk^d\frac{-2}{\sqrt{-g}}\frac{\delta I_2}{\delta g^{bd}}\\
= & -12k^bk_d\nabla _c\nabla _a\left(R^{[c}_{~~ebf}R^{d]fae}\right)\\
= & -6k^bk^d\nabla ^c\nabla _a\left(R_{cebf}R_d^{~fae}-R_{debf}R_c^{~fae}\right)\\
= & -6\nabla ^c\nabla _a\left(R_{cekf}R_k^{~fae}-\zeta ^2k_ck^a\right)\\
= & -6\nabla ^c\nabla ^a\left(R^e_{~cfk}R^{~~f}_{ea~k}\right)+6\partial _k^2\left(\zeta ^2\right)\\
= & -6\partial _k^2\left(R^e_{~lfk}R^{~~f}_{el~k}\right)+6\partial _k^2\left(\zeta ^2\right).
\end{aligned}
\end{equation}
The result is then
\begin{equation}
\Delta Q_{\text{ct2}}=T_{kk\text{, ct2}}-\frac{1}{2\pi }S_{\text{ct2}}''=0.
\end{equation}

For the final counter-term $I_1$, notice that $R_{akck}=\zeta k_ak_c$ implies $R^{ab}_{~~ck}R_{abdk}=\varsigma k_ck_d$. Contracting both sides with $l^cl^d$ gives the function $\varsigma  =R^{ab}_{~~lk}R_{ablk}$. Thus we find
\begin{equation}
R^{ab}_{~~ck}R_{abdk}=k_ck_dR^{ab}_{~~lk}R_{ablk}.
\end{equation}
This gives
\begin{equation}
\begin{aligned}
S'_{\text{ct1}}= & \partial _k\left(-12\pi R^{cdef}R_{ef}^{~~ab}g^\perp _{~[a|c|}g^\perp _{~b]d} \right)\\
= & \partial _k\left(-24\pi R^{ef}_{~~lk}R_{eflk}\right),
\end{aligned}
\end{equation}
\begin{equation}
\frac{1}{2\pi }S_{\text{ct1}}''=-12\partial _k^2\left(R^{ef}_{~~lk}R_{eflk}\right),
\end{equation}
and
\begin{equation}
\delta I_1=\int d^dx\sqrt{-g}3R^{cdef}R_{ef}^{~~ab}\left(-2\nabla _a\nabla _c\delta g_{bd}\right),
\end{equation}
\begin{equation}
\begin{aligned}
T_{kk\text{, ct1}}= & k^bk^d\frac{-2}{\sqrt{-g}}\frac{\delta I_1}{\delta g^{bd}}\\
= & -12k_bk_d\nabla _c\nabla _a\left(R^{cdef}R_{ef}^{~~ab}\right)\\
= & -12\nabla ^c\nabla ^a\left(R^{ef}_{~~ck}R_{efak}\right)\\
= & -12\partial _k^2\left(R^{ef}_{~~lk}R_{eflk}\right).
\end{aligned}
\end{equation}
The result is then once again that
\begin{equation}
\Delta Q_{\text{ct1}}=T_{kk\text{, ct1}}-\frac{1}{2\pi }S_{\text{ct1}}''=0.
\end{equation}

The above results are summarized in table \ref{tb:6dcounterterm}.

\bibliographystyle{utcaps}
\bibliography{all}

\end{document}